\documentclass[twocolumn]{aastex631}

\shorttitle{Modeling disk chemistry in moderate external UV fields}
\shortauthors{Gross et al.}

\usepackage{hyperref}
\usepackage{float}
\usepackage{subfigure}
\usepackage{graphicx} 
\usepackage{scalerel}
\usepackage{natbib}

\newcommand{\nthp}{N$_2$H$^+$}
\newcommand{\GO}{$G_0$}
\newcommand{\nt}{N$_2$}
\newcommand{\hh}{H$_2$}
\newcommand{\waterg}{H$_2$O(gr)}
\newcommand{\hcop}{HCO$^+$}

\newcommand{\cp}{C$^+$}

\newcommand{\cth}{C$_2$H}

\begin{document}

\title{Modeling the Impact of Moderate External UV Irradiation on Disk Chemistry}

\correspondingauthor{Rachel E. Gross}
\email{reg4ff@virginia.edu}

\author[0000-0002-0477-6047]{Rachel E. Gross}
\affiliation{University of Virginia, Department of Chemistry, Charlottesville, VA 22904, USA}
\author[0000-0003-2076-8001]{L. Ilsedore Cleeves}
\affiliation{University of Virginia, Department of Astronomy, Charlottesville, VA 22904, USA}
\affiliation{University of Virginia, Department of Chemistry, Charlottesville, VA 22904, USA}

\begin{abstract}
\noindent

The chemistry within a protoplanetary disk is greatly affected by external radiation from the local stellar environment. Previous work has focused on extreme radiation fields, representative of the center of something like the Orion Nebula Cluster. However, even in such environments, many disks exist at the edges of a cluster where the lower stellar density leads to radiation fields weaker by orders of magnitude compared to the center. We present new chemical models of a T-Tauri disk in the presence of a moderately increased interstellar radiation field (ISRF). Such an environment has a background UV strength of 10 to 100 times higher than the galactic average ISRF. Moderate radiation fields are among the most prevalent disk-harboring environments and have interesting implications for the chemistry of the outer disk radii. We find that the external UV radiation creates an outer ionization front that impacts the cold disk chemistry to varying degrees, depending on outer disk structure. Certain molecules like \cp, \nthp, C, and CS are more strongly impacted by the ISRF in their abundance, column density, and observable emission. Other abundant species like \hcop\ and CO are less affected by the external UV flux in the outer disk under such moderate UV conditions. Further, we demonstrate that the chemistry occurring in the inner tens of au is relatively unchanged, which suggests that even in moderately externally irradiated disks, the inner disk chemistry may be more similar to isolated disks like those in, e.g., the Taurus and Lupus star-forming regions. 
\end{abstract}


\section{Introduction} 
\label{sec:intro}

Most young stars form in clusters with high stellar density, and their circumstellar disks are subject to the enhanced external UV irradiation from nearby massive stars \citep{lada2003, adams2010,parker2020}. Enhanced far-ultraviolet (FUV) radiation in dense cluster environments can drive ionization in the surface layers of the disk, changing and shaping the disk's molecular composition \citep{walsh2013}. A dense, star-forming region, like the Orion Nebula Cluster (ONC) which contains massive O and B type stars, will have a much higher interstellar radiation field (ISRF) compared to isolated or quiescent star forming regions, such as Taurus or Lupus, where the lack of nearby massive stars results in a background radiation field that is closer to the galactic average, defined as $G_0=1$ \citep{habing1968}, such as the case of IM Lup, which is consistent with $G_0\sim4-8$ \citep{cleeves2016}. Most planetary systems form in disks encircling low-mass stars who are members of densely populated cluster environments similar to the ONC, and these disks appear to be chemically distinct from those residing in lower mass regions \citep{mann2014,eisner2018,Boyden2023}. Our Sun and Solar System are also thought to have formed in such a clustered environment \citep{adams2010}. Thus, studying disk evolution in these dense clusters is essential to understanding how an elevated UV environment ($G_0>>10$) affects the chemistry in planet-forming disks and ultimately the origins of our own planetary system.

However, our knowledge of disk chemistry in clusters has been relatively limited. At 400~pc, the ONC is the closest dense star forming region. Between the distance and potential cloud contamination from the nebula, it can be difficult to directly observe the chemistry taking place  \citep[e.g.,][]{williams2014,bally2015}. \cite{boyden2020} identified twenty three gas disks in the ONC using the bright lines of CO $J=3-2$ and \hcop\ $J=4-3$, which constitutes the largest sample of disk gas detections toward a relatively dense cluster to date. Large samples have been further limited by the fact that the harsh UV environment of the cluster can play a role in disks' physical disruption \citep[e.g.,][]{walsh2013,mann2014}. Radiation from massive O and B type stars can drive extreme mass loss from photoevaporative winds \citep[e.g.,][]{johnstone1998}. Theoretical studies also find that destruction timescales for photoevaporated disks can be quick, within 1~Myr, and these disks are effectively depleted of their planet-forming material \citep{adams2004,haworth2018a}. Despite the destructive effects of the OB stars, \cite{Boyden2023} found that at least some ONC disks retain molecular abundances more like the ISM compared to counterparts in isolation, which seem to be ``volatile poor.'' It was speculated that this effect may arise in part due to their warmer outer disk temperatures resulting in less processing into larger molecular species. The rich chemistry of mildly irradiated disks was recently confirmed by \citet{Diaz-Berrios2024}, who reported H$_2$CO, HCN, and C$_2$H in two proplyd systems with similar emission properties as more isolated disks.

To date, most chemical models of irradiated disks have focused on the extremes, either an isolated disk with a quiescent background radiation field at the galactic average, or that of a cluster with external radiation at thousands if not tens of thousands of times elevated above the background \citep[e.g.,][]{walsh2013, Boyden2023}. There are a significant number of disks that reside at intermediate distances from the cluster center and experience a substantially lower external UV field than the disks in the inner cluster, closer to G$_0$ values of $\sim100$. Such disks are expected to retain more of their gas \citep[due to reduced photoevaporation][]{haworth17} and may experience unique photochemistry. These intermediate UV fields are also most commonly present in star-forming regions across our Galaxy \citep[e.g.,][]{fatuzzo2008,parker2021}. 

We present a systematic analysis of the UV-driven chemistry occurring within disks at intermediate external radiation fields, representative of the edge of a young stellar cluster like the ONC. We aim to determine if the environment still plays a strong impact at such locations as has been found in simulations of chemistry at higher G$_0$. Our analysis focuses on commonly observed gas-phase species, and we explore how these species are changed by the strength of the UV field and the structure of the disk. We then use these results to provide guidance for how sub-millimeter observations of molecules in disks can be used to constrain the degree of external radiation a given system experiences.

\section{Methods}\label{sec:methods}

\subsection{Physical Model}\label{sec:phys}

The structure of the disk will play an important role in determining how much radiation from both the central star and external sources influences the disk's composition. For example, at fixed disk mass, a more extended outer radius allows for greater penetration of both stellar and interstellar UV photons, causing molecules to desorb and dissociate, which alters the chemistry \citep{walsh2013}. The ONC is known to contain a variety of disk morphologies, where the cluster center tends to have more compact and optically thick disks compared to disks in lower mass star-forming regions \citep{eisner2018}. Further out, $1-2$~pc away from the ONC center, more extended disks (with sizes as large as 1000~ au) are seen \citep[e.g.,][]{miotello2012}.

To investigate the impact of the density structure versus the role of external UV strength, we consider models with different outer radii for the same total disk mass. Three generic disk models were adopted from \citet{anderson2021} and are briefly described here. All three disks encompass a 1 $M_{\odot}$ T Tauri star with a radius of 2.8~$R_{\odot}$ and an effective temperature of 4300~K. The fiducial disk models are azimuthally symmetric and have a total gas mass of 0.01 M\textsubscript{\(\odot\)}. The inner radius for all three models is set to 1~au.  The surface density profile of the disk is described by the standard \citet{lyndenbell1974} formalism of a viscous disk, 
\begin{equation}
\Sigma_{gas}(R) = \Sigma_{c}(R/R_{c})^{-1}\exp[-(R/R_{c})].
\end{equation}
The values for the characteristic surface density, $\Sigma_{c}$, and characteristic radius, $R_c$, for each of the three disk models are shown in Table \ref{modelparams}.  The characteristic radius essentially sets the radial scale at which the disk goes from having the surface density dominated by a power-law to an exponential, which tapers the disk. In practice, as the disk radius is varied, the density of the disk changes due to the same amount of material in the disk, just extended or compacted depending on the disk radius. The three disk models have an outer radius of 50~au, 100~au, and 200~au, where the outer disk is effectively ``truncated'' at this radius. In the calculation of the external radiation field described in Section~\ref{sec:isrf}, this outer radius is considered the disk edge beyond which the gas density is zero. For each outer radius, the characteristic radius is scaled such that the disk model largely falls within the outer extent, see Table~\ref{modelparams}.

\begin{table}[htbp]
\centering
\caption{Disk Geometry Parameters}
\begin{tabular}{l c c c }
\hline 
\hline 
Model Parameter & Model 1 & Model 2 & Model 3 \\
        \hline
 $R_{\rm outer}$ (au)   &50  &100 &200     \\
 $R_{c}$ (au)    &15   &30  &60    \\
    $\Sigma_{c}$ at $R_{c}$ (g~cm$^{-2}$)    &69.64  &16.75  &4.2   
    \end{tabular}
    \label{modelparams}
\end{table}

We assume the dust mass is globally 1\% of the gas mass. For calculating the dust temperature structure, we use the TORUS code \citep{harries2004}, which calculates dust temperatures assuming the disk is passively irradiated (no accretion heating). For the properties of the dust, we assume the dust is made up of a large and a small population, where each population contains a range of dust grain sizes with a different maximum size. The large dust population has a maximum size of 1~mm and contains 90\% of total dust mass in total, and the small population has a maximum size of 1~$\mu$m. Both populations independently follow an MRN distribution (power law of -3.5) in grain abundance versus size. The large dust grains are settled to the midplane with a scale height of 10~au at a radius of 100~au.

To compute the UV and X-ray fluxes arising from the central star, we use the Monte Carlo radiative transfer code from \cite{bethell2011u}. For the UV, we use the same spatially dependent opacities as were used for the dust temperature calculations in TORUS. The wavelength-integrated stellar UV flux is $F_{\rm UV} = 5.129\times10^{10}$~photons cm$^{-2}$ s$^{-1}$. For the X-rays, the local opacity is based on each positions' gas-to-dust ratio, which sets the cross section for absorption based on \citet{bethell2011x}. Scattering is computing following \citet{cleeves2016org}. The input stellar X-ray spectrum is between 1 and 20 keV and contains a total luminosity of $10^{30}$~erg~s$^{-1}$. The cosmic ray ionization in all the models is the solar system minimum model from \citet{cleeves2015}.

Besides driving photochemistry, the UV is also an important component in setting the disk gas temperatures via the local photoelectric heating effect on the gas \citep{gorti2009}. Both the UV field from the star as well as the ISRF (see Section~\ref{sec:isrf}) are combined to estimate the additional UV heating that the gas experiences, and are based on fits to thermo-chemical models originally made by \citet{bruderer2013}. The heating and cooling balance is solved to determine gas decoupling from the dust in the upper layers of the disk atmosphere where gas temperatures can exceed the dust temperatures.

\subsection{Chemical Model}

The chemical abundances were determined by using the 2D time dependent gas-grain chemical model from \citet{fogel2011} as updated in \citet{cleeves2014}, \citet{cleeves2016}, and \citet{anderson2021}. This model utilizes the rate equation method to model the chemical evolution as a function of time, and includes 644 chemical species and 5944 chemical and physical processes.  
The reactions included in the model are ion-neutral, neutral-neutral, ion dissociative recombination, photon-driven chemistry, freeze-out of molecules, thermal and non-thermal desorption, and the vertical self-shielding of H$_2$, CO, and \nt\ as described in \citet{fogel2011} and \citet{cleeves2016}. 
In the case of CO, this molecule efficiently shields at a column density of $\sim$ 10$^{15}$ cm$^{-2}$, decreasing its photodissociation rate. CO can also be mutually shielded by atomic and molecular hydrogen, which is included in our models from \citet{visser2009} and \citet{Lee1996}. Self-shielding is especially important since these three species are highly abundant and play a major role in shaping how radiation can interact with the different compositional layers in the disk \cite{visser2009}. The initial chemical composition aims to reflect a typical molecular cloud (see Table \ref{abundances}) and the chemical calculations are evolved for 1 Myr, at which point the time-evolving abundances have approximately leveled out to a ``pseudo'' steady-state.

\begin{figure*}
    \centering
    \includegraphics[scale=0.4]{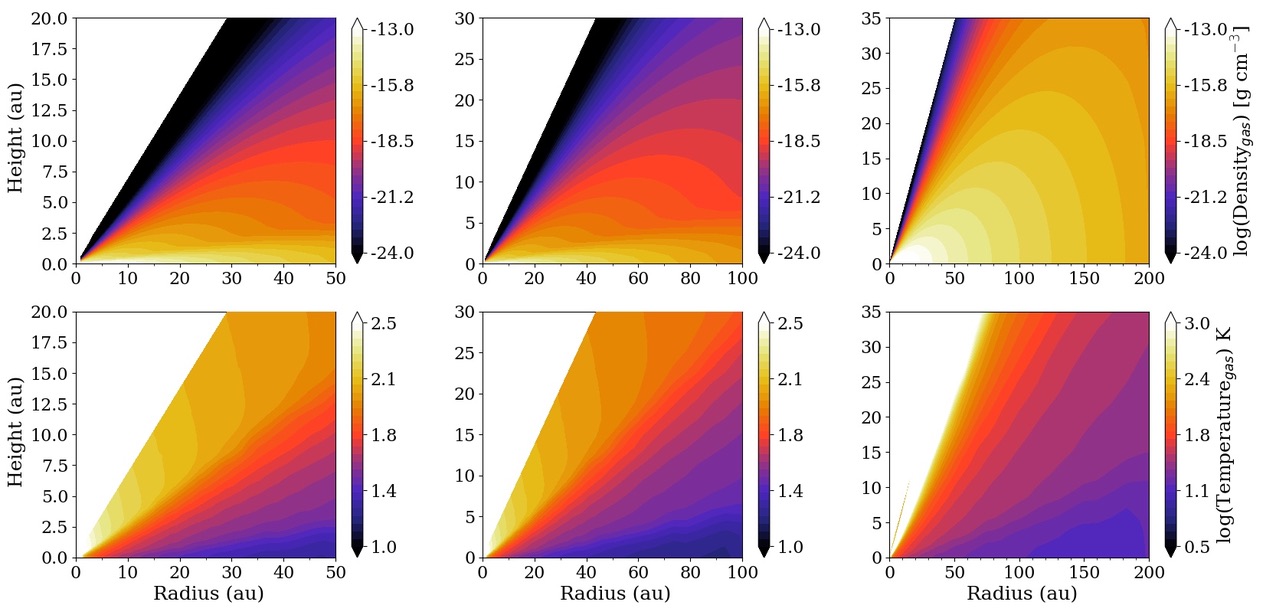}
    \caption{Gas density and temperature as a function of height and radius in the disk models. Top row: Gas density for the 50~au, 100~au, and 200~au models. Bottom row: Gas temperature for the three models. The more extended disk is warmer due to the overall lower density.} 
    
    \label{rho_temp}
\end{figure*}

\begin{table}[h]
\caption{Initial Abundances relative to Hydrogen}

\noindent\makebox[0.38\textwidth]{
\begin{tabular}{lr|lr} 
\toprule
\hh			&	$5.0 \times 10^{-1}$		&	Grain		&	$6.0 \times 10^{-12}$	\\
\waterg		&	$8.0 \times 10^{-5}$		&	CO			&	$9.9 \times 10^{-5}$		\\
O			&	$1.0 \times 10^{-8}$		&	C			&	$5.0 \times 10^{-9}$		\\
O$_2$		& 	$1.0 \times 10^{-8}$		&	NH$_3$		&	$4.8 \times 10^{-6}$		\\
He			&	$1.4 \times 10^{-1}$		&	HCN			&	$1.0 \times 10^{-8}$		\\
N$_2$		&	$3.5 \times 10^{-5}$	&	C$^+$		&	$1.0 \times 10^{-9}$		\\
CN			&	$6.0 \times 10^{-8}$		&	HCO$^+$		&	$9.0 \times 10^{-9}$		\\
H$_3$$^+$	&	$1.0 \times 10^{-8}$		&	C$_2$H		&	$8.0 \times 10^{-9}$		\\
S$^+$		&	$1.0 \times 10^{-11}$	&	CS			&	$4.0 \times 10^{-9}$		\\
Si$^+$		&	$1.0 \times 10^{-11}$	&	SO			& 	$5.0 \times 10^{-9}$		\\
Mg$^+$		& 	$1.0 \times 10^{-11}$	& 	Fe$^+$		&	$1.0 \times 10^{-11}$	\\

\end{tabular}} \label{abundances}
\end{table}


\subsection{External UV Model}\label{sec:isrf}

While the cluster environment can influence disk evolution in many different ways, we focus our analysis on the role of external UV since this component will be the most elevated in the cluster environment for the duration of the cluster lifetime \citep[e.g.,][]{adams2010, haworth2017, winter2018}. In this section, we describe our treatment of the radiation transfer of the external radiation into the disk. We assume the ISRF is not time-evolving throughout our simulations to investigate what magnitude causes clear chemical changes. In reality, the changing stellar population in the cluster or dynamical interactions between stars can change the local external radiation field to vary on short timescales; \citep{winter2019a,winter2019b} however, this is beyond the scope of the present work. We also assume the ISRF is isotropic, rather than coming from a single bright source or cluster of bright sources for simplicity. 

\begin{figure*}
    \centering
    \includegraphics[scale=0.3]{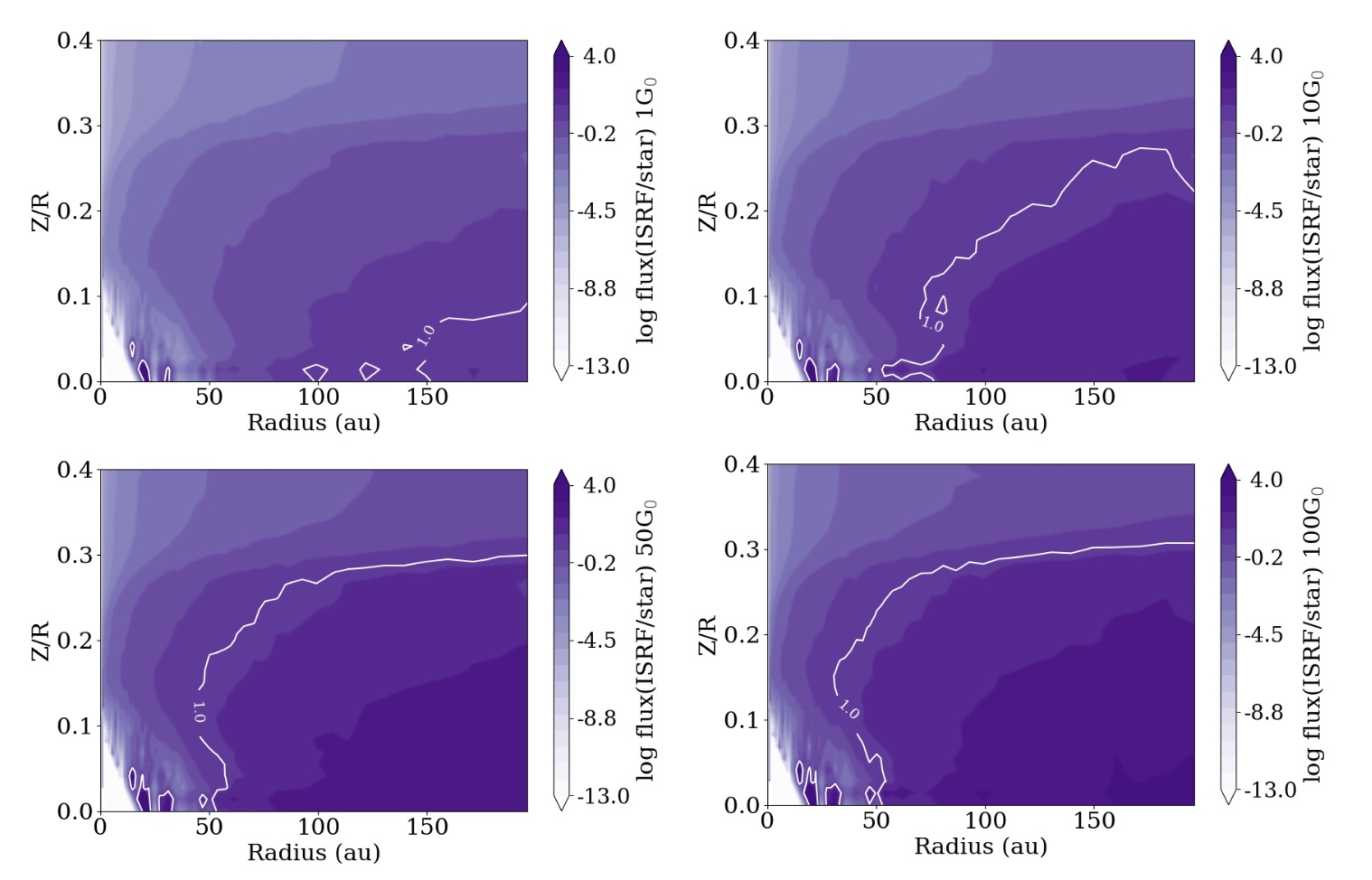}
    \caption{Comparison of the ISRF to UV from the central star as a function of radius and height, note the log scale. White contours indicate where they are equal. As the \GO\ increases, the ISRF increasingly dominates over the central star, moving the line inwards.} 
    
    \label{isrf_stellar}
\end{figure*}

Whether the ISRF dominates the stellar field or not depends on the strength of both UV sources as well as the density structure of the disk. The relatively lower densities of the outer disk combined with geometrical effects allow the ISRF to dominate the UV of the central star. The impact of the ISRF on the bulk disk chemistry is primarily limited by dust attenuation as well as self-shielding. A common approximation for estimating the dust extinction to the ISRF is to calculate the vertical column density of material from the disk surface to a given point. This can underestimate the extinction, especially when considering lines of sight along the midplane and so we instead use the method of \citet{cleeves_isrf} to estimate a 3D-averaged absorption optical depth to each location in the disk. The average is computed along each line of sight (``rays'') for 150 lines to a localized point in the disk, of which there are 150 radial points and 80 vertical points. These ``rays'' are evenly distributed over 4$\pi$ steradians to calculate the effective optical depth at each point. This optical depth is then applied to compute the extincted ISRF spectrum from 912 and 2000~\AA. The details of this approach are described further in \cite{cleeves_isrf}.  

\begin{figure*}
    \centering
    \includegraphics[scale=0.42]{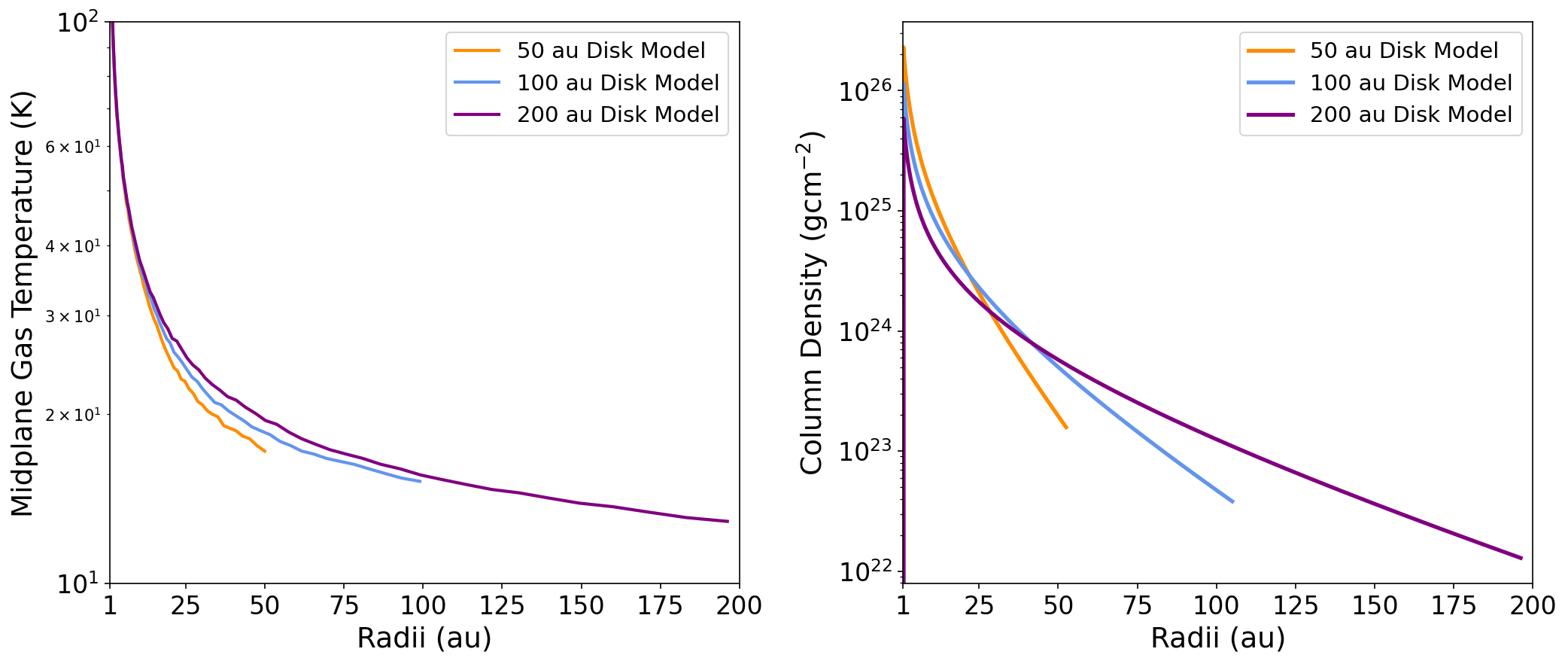}
    \caption{Right: The gas temperature profile at the midplane for each of the three disk physical models. Left: The total gas column density for all three disk models.} 
    \label{gas_temp_colden}
\end{figure*}

The average integrated strength of the UV background radiation field is defined by \GO\ \citep{habing} with a value of $\sim 1.6 \times 10^{-3}$ erg cm$^{-2}$ s$^{-1}$ over the wavelength range of 912 to 2400~\AA. 
For our models we chose four values of \GO, ranging from 1~\GO, which represents the external UV of the galactic average or for an isolated disk. Multiples of 10, 50 and 100~\GO\ are then used to simulate moderately enhanced external UV fields.  Figure~\ref{isrf_stellar} shows a comparison between the UV from the central star versus the environment for the 200 au radius model at the different \GO\ values.

As mentioned above, we approximate the gas heating from the photoelectric effect arising from the combination of the stellar UV, which includes contribution from Lyman-\(\alpha\) (Ly\(\alpha\)) radiation, and the interstellar radiation field (ISRF), which does not. Neutral hydrogen (HI) in the interstellar medium (ISM) efficiently absorbs and scatters Ly\(\alpha\) photons, significantly reducing their presence in the external UV input. Thus, the contribution of Ly\(\alpha\) radiation can be ignored in the external UV field for our chemical models. Instead, Ly\(\alpha\) radiation is primarily accounted for in the stellar UV component in the inner disk regions where the stellar flux is more dominant \citep[e.g.,][]{bergin2003, fogel2011}.

The UV contribution from the ISRF, \(F_{\text{ISRF}}(\lambda)\), is combined with the stellar UV flux, \(F_{\text{star}}(\lambda)\), in a wavelength-dependent manner and then folded into the chemical model’s photo-rates. The photoionization and photodissociation rates are based on the wavelength-dependent data provided in the van Dishoeck database\footnote{\url{http://www.strw.leidenuniv.nl/~ewine/photo/}} \citep[e.g.,][]{vanDishoeck2006, vanHemert2008, Heays2017}. The combined stellar and external UV fields are also considered to estimate the gas temperature \citep[based on models fit to the results of][]{bruderer2012}. The photodesorption yield varies per species and was last updated in Anderson et al. (2021), who incorporated values from the same Leiden photodissociation repository cited above. For species without specific values, a default yield of \(10^{-3}\) molecules/photon is assumed.

The gas temperature profiles for the midplane in the three disk models are shown in Figure~\ref{gas_temp_colden}. As the three physical disk models are constrained to the same total mass, the 50~au disk is quite compact and dense, while the 200~au disk is more extended by comparison. The midplane temperature of the 200~au disk is warmer than either the 100 and 50~au disks. The compact, 50~au disk is ultimately the coolest throughout the midplane, as it is the most shielded, only reaching \(\sim 20\,\text{K}\) at the outer edges of the disk. The total gas columm density for each of the three physical models is also shown in Fig \ref{gas_temp_colden}.

\section{Results and Discussion}\label{sec:results}

\subsection{ISRF effects on neutral species}\label{sec:resultsneutrals}

\begin{figure*}
    \centering
    \includegraphics[scale=0.19]{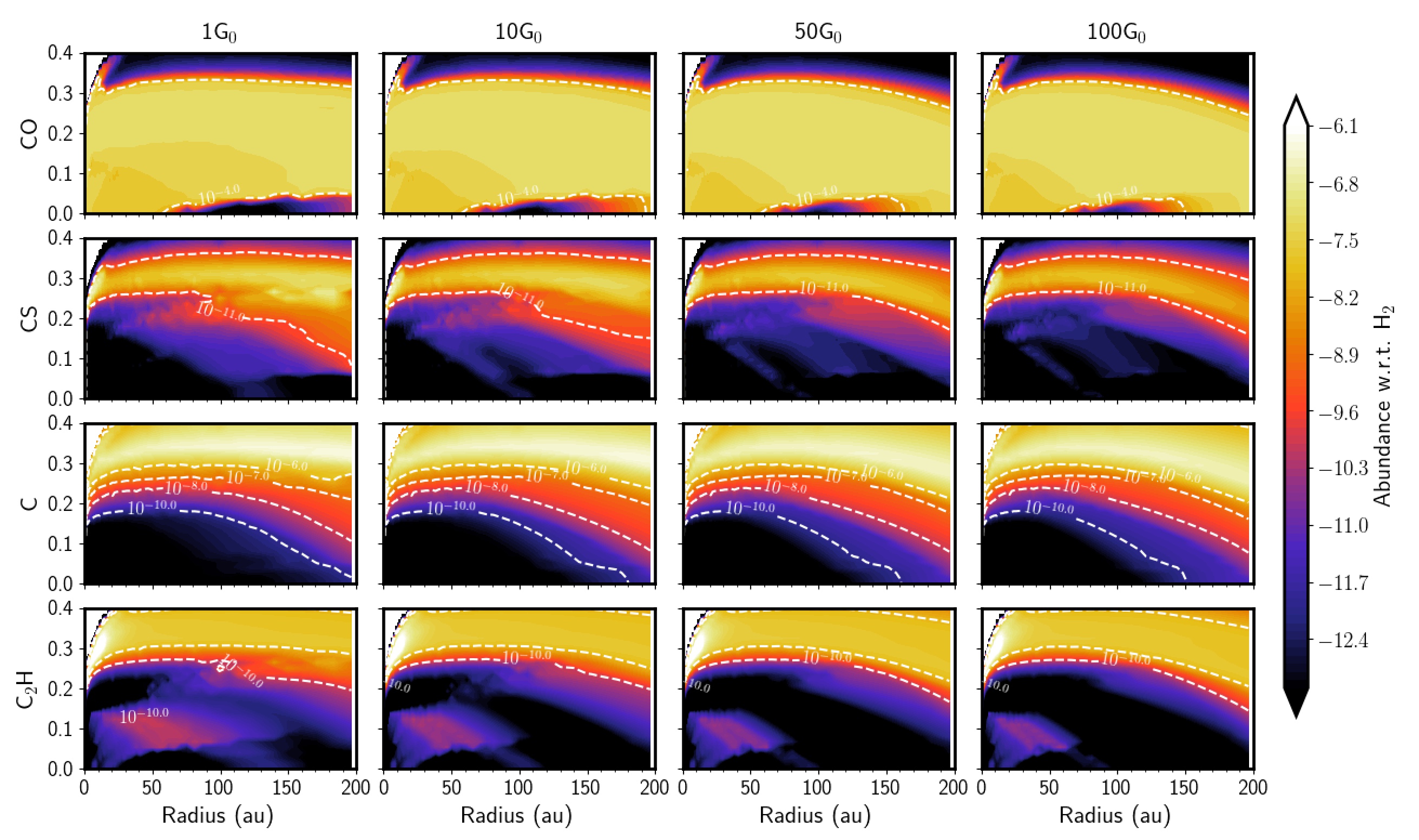}
\caption{Abundances with respect to hydrogen for a sample of neutral gas-phase tracers for the 200~au disk model. From top to bottom: CO, CS, and atomic C. From left to right, columns show models of increasing ISRF strength: 1~\GO, 10~\GO, 50~\GO, and 100~\GO.} 
    
    \label{abund_neutrals}
\end{figure*}

\begin{figure*}
    \centering
    \includegraphics[scale=0.19]{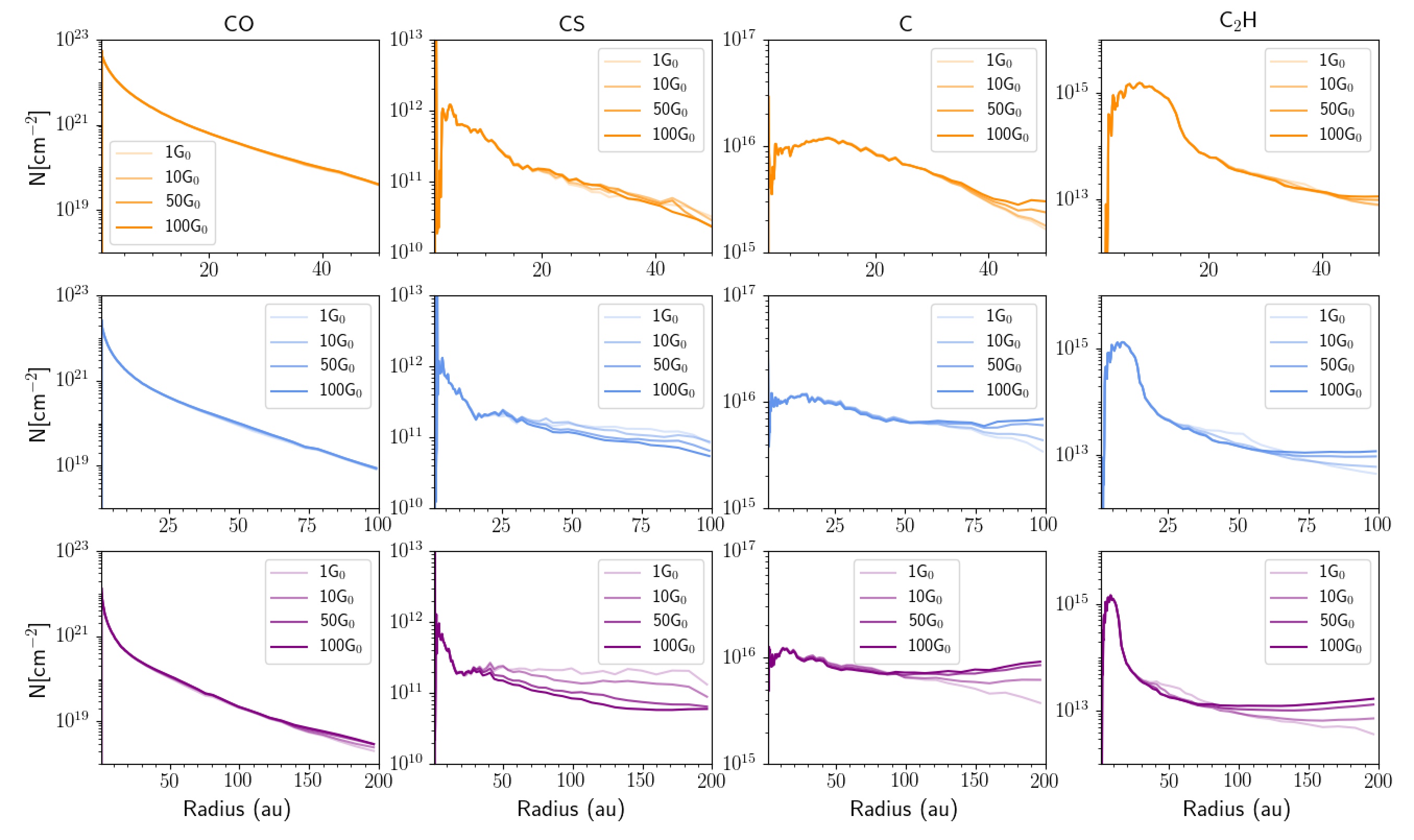}
    \caption{Column densities for CO, CS, C, and \cth\ for three disk models. The top shows the results for the 50 au model, middle for the 100~au model, and bottom for the 200~au model. For each panel, the faintest line corresponds to the model with a 1~\GO\ ISRF field and the darkest line is 100~\GO. Column densities are calculated for all heights and both sides of the disk in our models.} 
    
    \label{colden_neutrals}
\end{figure*}

\subsubsection{CO}

Gas-phase CO is one of the most abundant, sub-mm observable molecules and thus is commonly used to trace a variety of physical and chemical processes. We find that the CO abundance and column density are not strongly impacted under the moderate UV conditions. The top row of Figure~\ref{abund_neutrals} shows the CO abundance at increasingly higher external \GO. Over-plotted is an abundance contour of $10^{-4}$ per H, and provides a visual boundary for where CO is disappearing either due to freeze out or photo-dissociation. 

As the \GO\ increases, the abundance of gas-phase CO increases near the midplane, pushing the contour line inwards towards the central star. At this position, photodesorption becomes competitive with freeze-out, where the grains are still $<20$~K. This secondary snowline effect is described in \cite{cleeves2016} and can be further enhanced by processes like radial drift, making the outer edge of the disk more susceptible to UV interaction and increased CO desorption. While we do not account for radial drift in our models, the 200~au disk is more extended and therefore has lower opacity in the outer radii. This allows for increased penetration of UV photons, and creates the secondary CO snowline formation starting at 50~\GO.

This effect is only seen slightly in the 100 au model and not at all in the 50 au model, shown in Appendix \ref{abund_neutrals100} and \ref{abund_neutrals50}. This is due to the density structure of the disk. All three physical disk models were constrained to the same mass, and so as the radius decreases the disks become much denser and have a higher opacity, more efficiently shielding the midplane from stellar and external UV photons.

This effect is not as obvious in the CO column density for any of the models, where CO shows only the slightest increase in column density at 100~\GO\ for the 200~au sized model (see first column of Figure \ref{colden_neutrals}). 
At these locations, CO chemistry is dominated by freeze-out, photo-desorption, and photo-dissociation. The molecular layer as traced by CO is pushed down with increasing UV as seen in the 2D abundance plots in Figure~\ref{abund_neutrals}, but remains similar in total column density. In principle, differences may be able to be derived by measuring CO emitting height channel-by-channel \citep[e.g.,][]{law22} and comparing it to other species \citep{pc23}, but using the CO intensity alone will be challenging, especially given the typical optical depth of CO emission in disks.

\subsubsection{CS}

CS is a classic dense gas tracer and a useful diagnostic for constraining the UV effects between the inner and outer disk. It is also a dominant carrier of volatile sulfur in disks \citep{keyte2023} and has been detected in multiple disks in quiescent regions \citep[e.g.,][]{Booth2023}. The CS $J=7-6$ line has been spectrally resolved in a large ONC disk in the outskirts of Orion by \citet{mann2014} and further analyzed by \citet{factor2017}.

Figure \ref{colden_neutrals} shows the column density of the three disk models. The inner radius up to 40 au is unchanged in all the models. Between 40 and 50 au we start to see the column density drop in the 100~\GO\ model compared to the 1~\GO\ model. There is a successive decrease in the column density as the \GO\ value is increased. At the outer edge of the disk around 175 au is almost an order of magnitude decrease in the 100~\GO\ model from the 1~\GO\ model, and the column density ratio between the two models is $\sim2.8$.

The CS in the middle row of Figure~\ref{abund_neutrals} shows a shrinking abundance in the outer disk between a Z/R of 0.1 and 0.2 as the external UV field is increased from 1~\GO\ to 100. We also see the contour band at the midplane disappear between 110 and 150 au at 100~\GO, denoting a drop of an order of magnitude in abundance. Thus, CS should be a useful anti-correlated tracer of external UV in cases when the disk is not overly dense and thus shielded to the changing UV.  

\subsubsection{C}

Neutral carbon exists in the mid- to upper-layers of the disk, between the CO dissociation region and where carbon is ionized in the surface layers, following typical PDR-like chemistry \citep{tielens1985} and recently directly confirmed via resolved ALMA observations in a disk in the more quiescent Lupus star forming region \citep{law23}. 

Figure~\ref{abund_neutrals} shows the 2D abundance of neutral atomic C as a function of height and radius in the disk in varying external UV fields. In the inner radii, its abundance is very sensitive to UV photons arising from the central star, but the external UV contributes at the surface of the outermost radii.
For low UV, C sits in a  band at Z/R of $\sim0.3$. As the external UV flux increases, the band tapers downward, showing a higher abundance of C in the outer disk, as expected from increased CO photodissociation, and the C at the very top of the disk becomes photoionized into C$^+$. 

The column density of C shown in Figure~\ref{colden_neutrals} shows a trend of increasing column density when the external UV is enhanced. This effect can even be seen in the very outer radii of the 50 au and 100 au disk. Whereas the other chemical species show little change in the compact and dense 50 au disk, C shows a difference in column density ratio between 1~\GO\ and 100~\GO\ of $\sim$1.6. In the 200 au model, at the very edge of the disk, the column density of the 100\GO\ model is an order of magnitude higher and has a column density ratio of $\sim3$. The overall increase in column density suggests that photodissociation of CO out-competes photoionization of C, and this effect is stronger in the more extended lower-density disk where UV reaches further toward the disk midplane.

\subsubsection{\cth}

\cth\ abundance is dependent on the C/O ratio in the disk, as it forms through multiple pathways involving atomic C.
These carbon-rich environments often result from processes like the photodissociation of CO in enhanced UV environments, which increased available carbon for hydrocarbon chemistry \citep{bergin2016,bosman2021}.
In Fig \ref{abund_neutrals}, the abundance of \cth\ is shown in the bottom row, predominantly residing in two layers, with an upper layer constituting the majority of species and a smaller midplane layer within the disk. An increase in abundance is seen in the upper layers at heights between 0.2 and 0.25, as well as a decrease in the lower layer of \cth\ in the midplane region between 60 and 150~au.
The column densities of \cth\ in our models are consistent with predictions from disk chemical models that assume a C/O ratio between 0.4 and 1.0 in \citet{bergin2016}, demonstrating that \cth\ column densities are sensitive to the C/O ratio in the disk. \citet{bergin2016} demonstrated that even small increases in the C/O ratio can affect \cth\ column densities by up to four orders of magnitude. Fig \ref{colden_neutrals} shows the column density for \cth\ for each irradiated disk model. The trends follow closely to C and \cp, where we see an increase in column density in the outer disk radii especially for the 200~au disk model between 100 and 200~au.

\begin{figure*}
    \centering
    \includegraphics[scale=0.19]{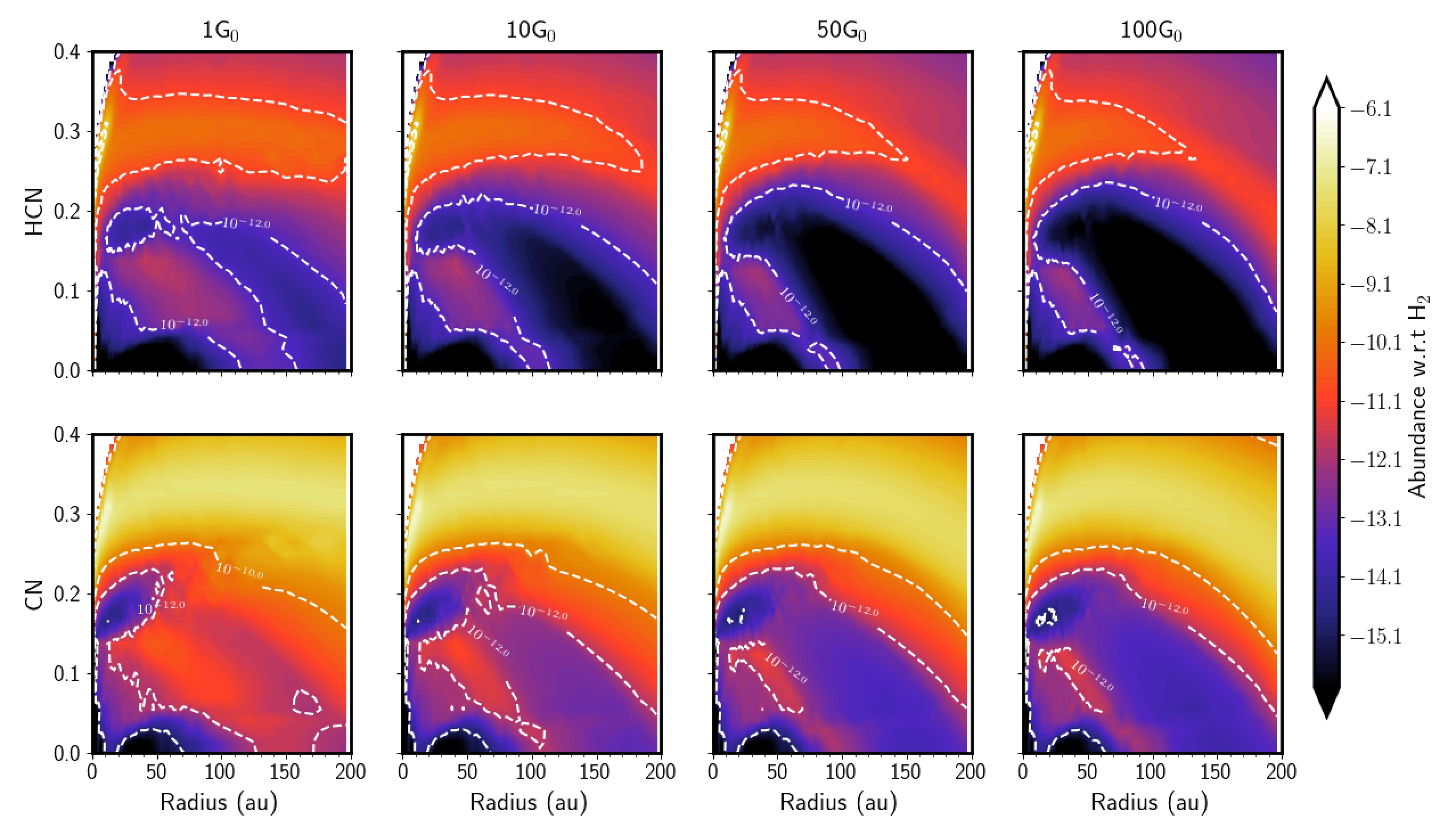}
    \caption{Abundances with respect to hydrogen for HCN and CN. From left to right, columns show models of increasing ISRF strength: 1~\GO, 10~\GO, 50~\GO, and 100~\GO.} 
    
    \label{hcn_cn_abund}
\end{figure*}

\begin{figure*}
    \centering
    \includegraphics[scale=0.42]{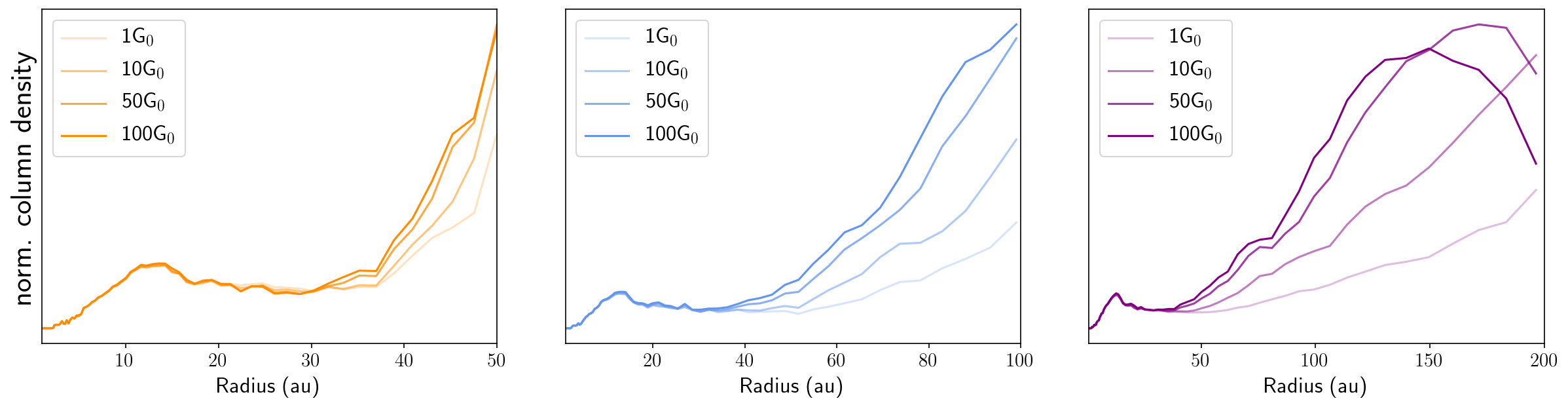}
    \caption{Column Density for the CN/HCN ratio for each of the \GO\ models. Note that as the external irradiation increases so does the ratio.}
    \label{cn_hcn_ratio}
\end{figure*}

\subsubsection{HCN \& CN}

The ratio of CN/HCN has been widely used as a tracer of UV irradiation in protoplanetary disks \citep{bergner2021}. HCN is photodissociated in regions of high UV, leading to an elevated CN/HCN ratio, particularly in the surface layers of the disk where stellar UV radiation is strongest. This trend is expected to hold true in environments subjected to external UV irradiation, such as the ONC. In deeper regions of the disk midplane, where UV flux is attenuated, a lower CN/HCN ratio is expected due to the reduced photodissociation of HCN by UV photons.
Our modeled CN/HCN column densities (Figure \ref{cn_hcn_ratio}) demonstrates this behavior, where increasing external UV irradiation results in a decrease in HCN abundance, particularly beyond 50~au when the external irradiation reaches 100~\GO. This results in an increase in the CN/HCN ratio. Figure \ref{hcn_cn_abund} shows that both HCN and CN are depleted in the outer disk midplane and molecular layers under higher UV flux conditions, with the HCN abundance decreasing by a few orders of magnitude. This suggests that external UV irradiation has a pronounced effect on the chemical composition of the outer disk, destroying both HCN and CN. It is important to note that our chemical model does not include vibrationally excited \hh, a potentially important formation pathway for HCN, which might mitigate the observed destruction by facilitating additional HCN production through reactions with CN.
The findings from \citet{walsh2013} provide a comparison, as they modeled a protoplanetary disk exposed to extreme UV irradiation from a nearby O-type star. Their study similarly shows that the CN/HCN ratio increases in the irradiated disk model compared to an isolated disk. \citet{walsh2013} also predicts that this elevated ratio can serve as a diagnostic tool for assessing the impact of external UV fields, a conclusion that aligns well with our results. However, they note that shielding in the midplane can allow for some molecular retention, even under extreme irradiation, whereas our models show a more pronounced depletion of both HCN and CN under moderate UV conditions.

\subsection{ISRF effects on ions}\label{sec:resultsions}

\begin{figure*}
    \centering
    \includegraphics[scale=0.185]{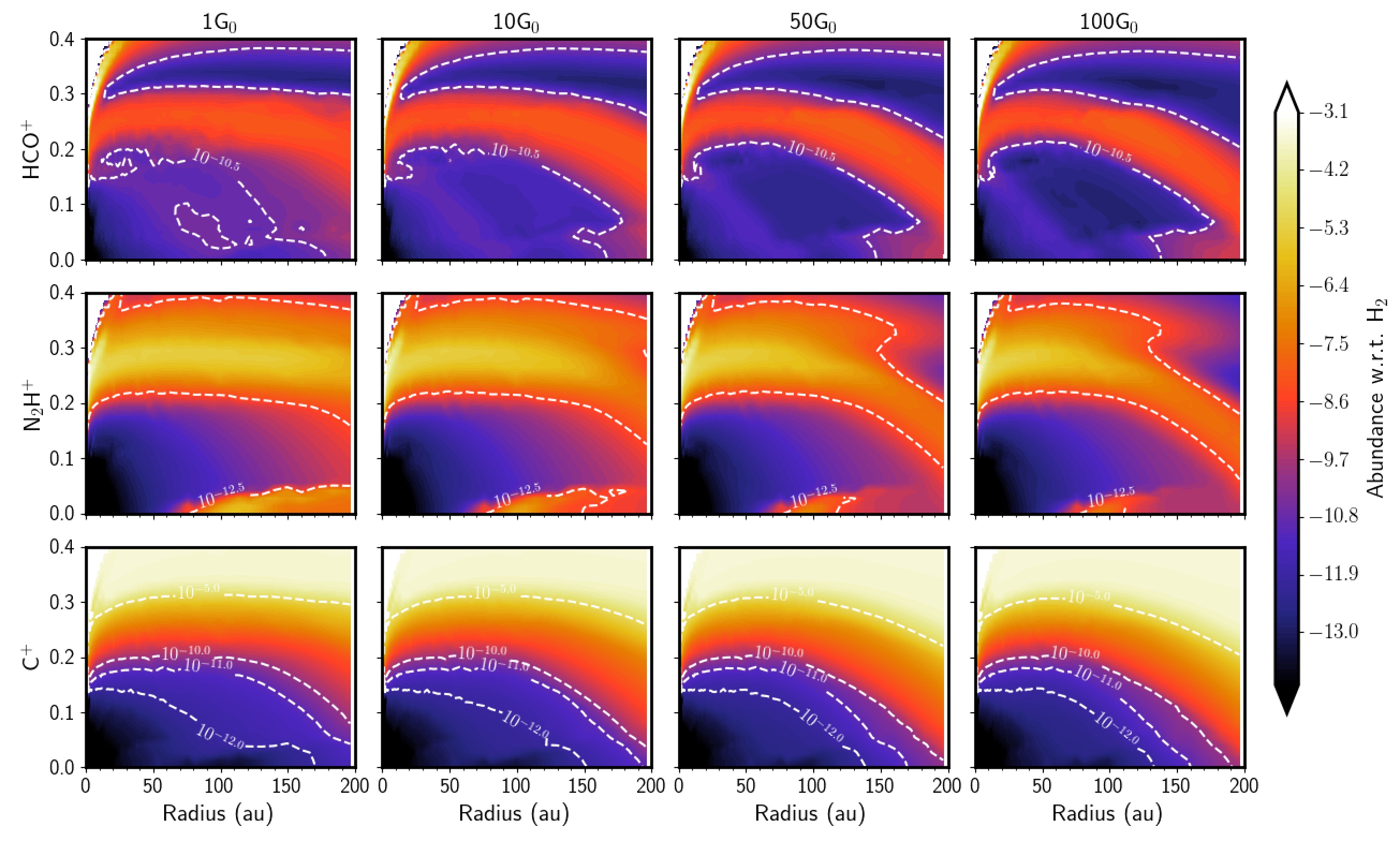}
    \caption{2D contour abundances with respect to hydrogen for a sample of neutral gas-phase tracers. Top row is \hcop: middle row is \nthp: and bottom row is \cp\ for the 200 au disk model. The first column is 1\GO, second is 10\GO, third is 50\GO, and last column is 100\GO. The white contours depict various abundances to trace the effects of external UV and are labeled.} 
    
    \label{abund_ions}
\end{figure*}

\begin{figure*}
    \centering
    \includegraphics[scale=0.19]{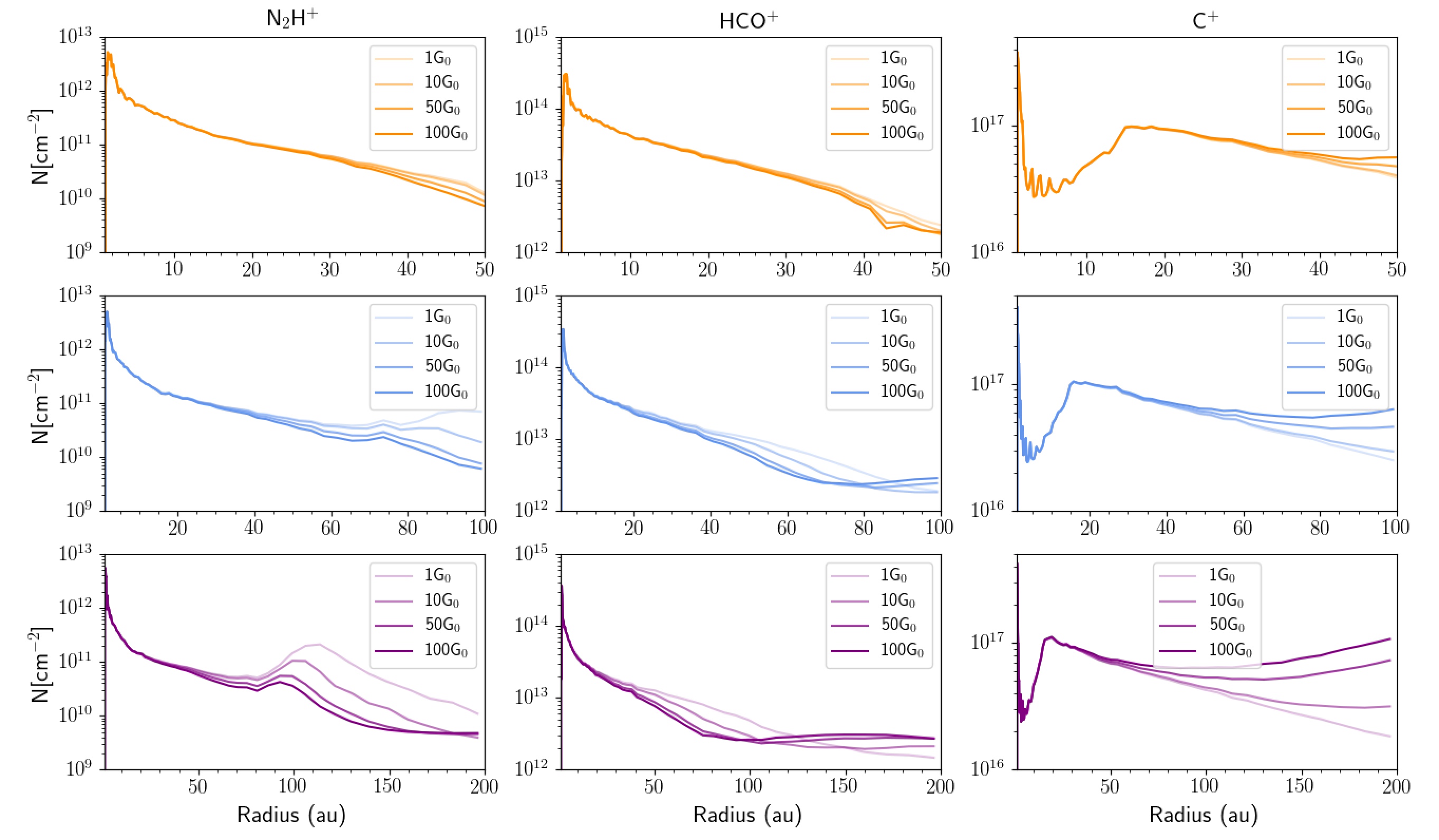}
    \caption{Column densities for \hcop, \nthp, and \cp\ for three disk models at all four \GO\ values as indicated in the legend.} 
    
    \label{colden_ions}
\end{figure*}

\subsubsection{\hcop}

\hcop\ is a common dense gas tracer and is a particularly useful ionization tracer within disks \citep[e.g.,][]{walsh12,cleeves15,seifert21,aikawa21,long24}. It is one of few species to be reliably detected in irradiated disks due to its bright emission lines \citep{Boyden2023}. \hcop\ has been found previously to be sensitive to both UV and X-ray radiation fields in disks without strong background radiation fields \citep[e.g.,][]{seifert2021} and so here we examine how the cluster environment alters this picture.

The column density of \hcop\ shows interesting features in Figure~\ref{colden_ions}. Initially, the column density decreases with increased irradiation between 25 and 100~au. This can be due to electron recombination destruction of \hcop\ which is elevated slightly in the 100~\GO\ model. In the 1~\GO\ model, the reaction rate of \nthp\ and CO is elevated in this same radii region due to more \nthp\ available. 

The further out in the disk, from 100 to 200~au, the trend switches to the irradiated column densities increasing. The rate by which \hcop\ is formed in the molecular layer of the outer disk between 100 and 200~au is increased by two orders of magnitude for the 100~\GO\ model. This main formation reaction is not dominated by the \nthp\ and CO reaction, despite there being more photodesorbed CO available in this region of the disk. Instead, there is an increased reaction of CO and H$_3^+$ in the outer disk of the irradiated model, forming \hcop\ at a higher rate.

\subsubsection{\nthp}

\nthp\ is a versatile molecule that has traditionally been used to trace disk chemistry closer to the midplane, especially the role of ionization from cosmic rays \citep{seifert2021,long24}. The precursor of \nthp, \nt, remains in the gas phase at much lower temperatures than other molecules \citep{oberg2005,vanthoff2017}, making it an ideal tracer of cold gas in the outer disk. \nthp\ is also used to put constraints on the CO snowline location in disks, as their chemistry opposes one another \citep[e.g.,][]{qi13,qi2019}. When CO is in the gas phase, it readily destroys \nthp, and so we can observe \nthp\ when CO freeze-out occurs. Thus, the combination of its dependence on CO chemistry and its relationship with the ionization of the disk gas make it an interesting case study for externally irradiated disk chemistry. 

Further studies by \citet{Trapman2022} demonstrate how the combination of \nthp\ and CO can provide a more accurate measurement of the gas mass in protoplanetary disks, correcting for the underabundance of CO due to processes like chemical conversion and freeze-out. This approach improves the accuracy of gas mass estimates by a factor of $5-10$, as demonstrated in more isolated systems like TW Hya, GM Aur, and DM Tau. \nthp and CO could therefore offer a reliable method to characterize the volatile budget available for planet formation \citep{Trapman2022}. More studies on irradiated systems would be needed to understand how this technique would work in different environments.

We find two bands of abundant \nthp\ in our models; one higher up in the ionizing layers of the disk and one near the midplane (see Figure~\ref{abund_ions}). This morphology is similar to what has been seen previously in models, e.g., \citet{vanthoff2017}. Focusing first on the higher layer at a Z/R of $\sim0.3$, CO photodissocation rates are relatively high such that the formation of \nthp\ can compete with its destruction by CO. As CO dissociation increases with increasing \GO\, this boundary zone drops to lower Z/R. At the same time, the increased photo-desorption of CO in the colder layers begins to destroy \nthp\, resulting in its net destruction with higher \GO.   
The midplane abundance of \nthp\ also begins to retreat simultaneously due to the same increased CO desorption. In the 200~au outer radius model, we see an order of magnitude decrease in the \nthp\ abundance in the outer radii between $\sim$120-200~au. This can also be seen in the same location of the \nthp\ column density shown in purple in Figure~\ref{colden_ions}. At 200~au the column density ratio between the 1~\GO\ and 100~\GO\ case is approximately a factor of 5 lower, demonstrating the effects of the UV irradiation. Even the increase from 1~\GO\ to 10~\GO\ shows a decrease in the column density outside of $R>100$~au.  \citet{walsh2013} finds a similar trend with \GO, albeit for a much higher UV flux. 

The compact disks are again less impacted by increasing \GO\ compared to the 200~au disk. Nonetheless, the 100~au disk model still shows a substantial decrease in outer disk column density, decreasing by an order of magnitude at the very outer edge of the disk going from 1 \GO\ to 100 \GO. The 50~au model, however, is minimally changed. Clearly, as the disk becomes more dense and the UV attenuation to the ISRF increases, there is less alteration of the \nthp, about a factor of $\sim2$ decrease beyond 40~au. 

In the 100~au and 200~au disk models, there is a local maximum in the \nthp\ column density that appears at 75~au and $\sim110$~au, respectively, shifting slightly with \GO. Similar ring-like features have been observed in disks in the literature \citep[e.g.,][]{qi2019} and have been associated with the CO snowline. In our model we can directly locate the midplane CO snowline at 45, 50, and 60~au for the models with outer radii of 50, 100, and 200~au, respectively. The increase in snowline position is expected since as the disk density drops, the central star can more efficiently heat the disk interior, pushing the CO snow-line outward. However, as can be seen in the 200~au model, the peak position moves inward by $10-20$~au with increasing \GO\ because of the increased electron recombination from the ionization of species like C in the outer disk. As the external UV field increases, the amount of free electrons in the disk also increases as other species are ionized as well. These electrons can increase the rate of electron recombination with \nthp\ by two orders of magnitude in the disk, especially for the 200~au disk which is less dense and much more penetrable by photons. 

Thus the utility of \nthp\ as a snowline tracer will be made more complex by interactions with the interstellar radiation field. For the \GO=1 case, the peak in \nthp\ is $1.8-1.9\times$ further out than the midplane snowline, in part due to a column density effect, but as \GO\ increases, the peak is only $1.6\times$ the midplane CO snow line distance. The peak location is impacted by the changing destruction mechanisms -- being dominated by CO (low \GO\ case vs. electrons and/or charged grains in the high \GO\ case). The higher electron density in the outer disk preferentially suppresses the outer edge of the \nthp\ abundance, shifting the local peak appearance inward, even if the dust temperature itself is not changing due to the changing ISRF.

\subsubsection{\cp}

The \cp\ abundance shows similar trends to the neutral C abundance for the 200~au disk, peaking just above the C layer as expected. In the upper layers, which dominate the \cp\ globally, the abundance can reach values as high as $\sim10^{-5}$ per H in the high \GO\ case. The midplane abundance of \cp\ is also impacted by the ISRF. For example, the midplane \cp\ abundance increases by an order of magnitude in the midplane around 150~au in the 200~au radius model between low and high UV.

The outer disk column density of \cp\ generally increases for all three models as \GO\ increases. The enhanced volatile C on the surface and the outer disk from the photodesorbed CO is easily ionized, as the first ionization potential of carbon is $\sim11.2$ eV. The column density ratio increases by a factor of 5 from 1~\GO\ to 100~\GO\ in the 200~au radius model.  While outer disk emission from \cp\ is challenging to observationally detect in disks, it is a key ion in driving the overall cation chemistry and is a useful diagnostic for understanding the broader chemical impact on related species like CO and \nthp. 

\subsection{Impact on Line Emission}\label{sec:LIMEsim}

\begin{figure*}
    \centering
    \includegraphics[width=\textwidth]{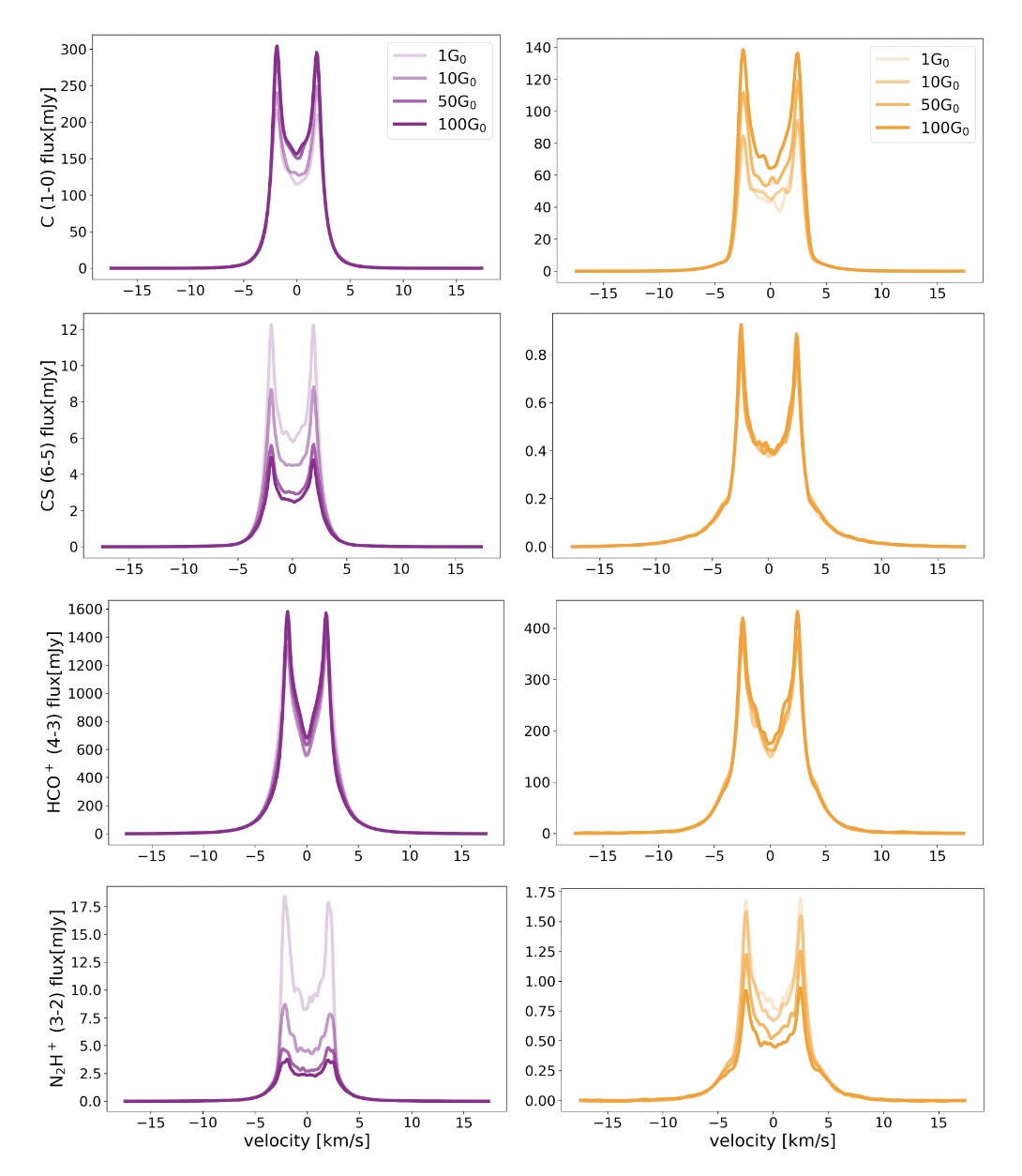}
    \caption{LIME radiative transfer models for select lines of CS, C, \hcop, and \nthp\ for the 200~au (purple) and 50~au (orange) disk models at varying \GO\ values. On average, the more extended 200~au disk case shows more affect in the flux from an enhanced external UV field. Note, the modeled disk inclination is 60\textdegree, and all of the models presented assume a default distance of 100~pc.}
    \label{spectra}
\end{figure*}

\begin{figure*}
    \figurenum{\ref{spectra}} 
    \centering
    \includegraphics[width=\textwidth]{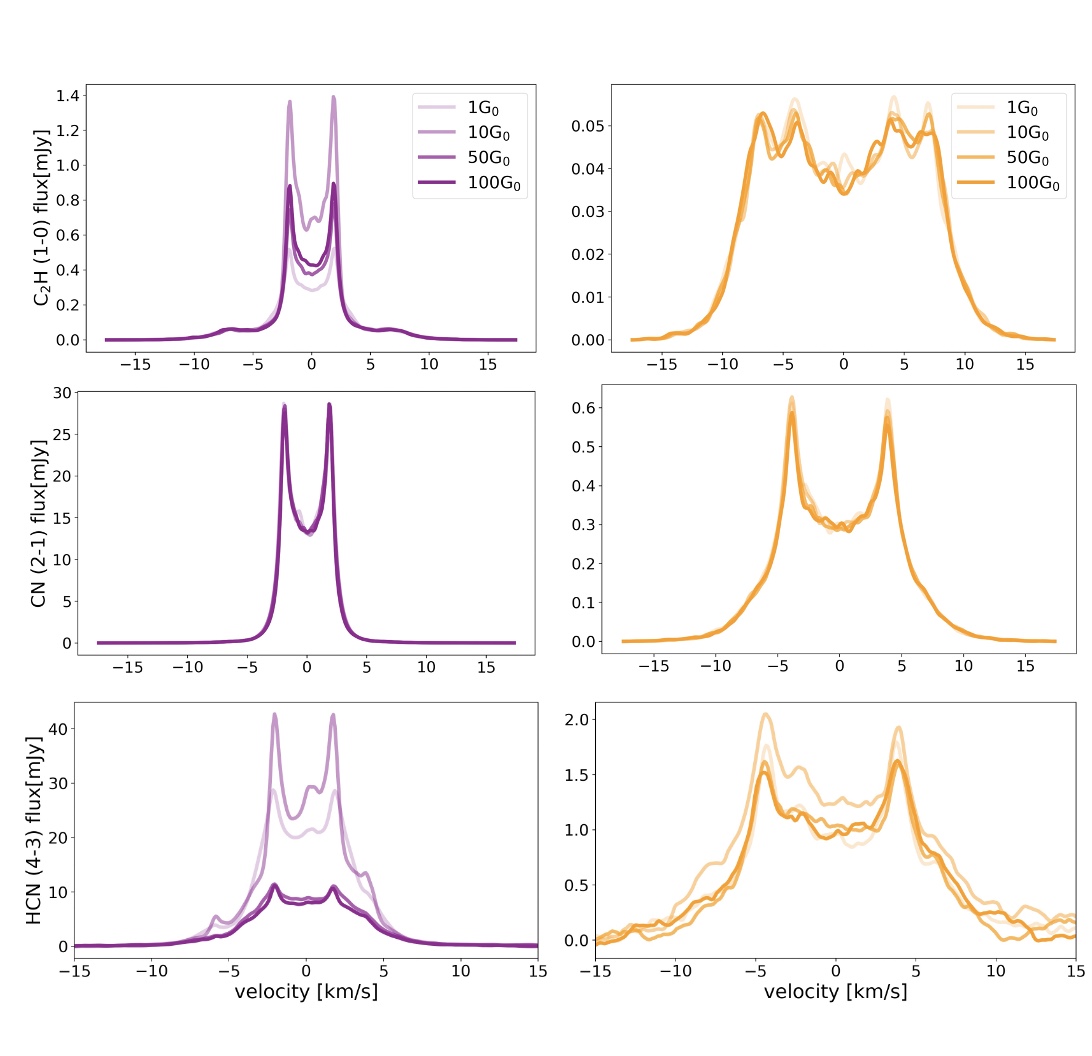}
    \caption{Fig cont'd for LIME radiative transfer models for select lines of \cth, CN, and HCN for the 50~au and 200~au disk models at varying \GO\ values. The purple plots are the 200~au models and the orange correspond to the 50~au models. Note the \cth\ and HCN spectra include hyperfine components,  and small fluctuations are due to random noise in the radiative transfer.}
\end{figure*}

\begin{table*}
\centering
\caption{Total number of particles and integrated flux values for species in Fig 10 for all four \GO\ values at 50 and 200~au models at 100~pc distance.}
\begin{tabular}{lcccccccc}
\hline
Molecule  & \multicolumn{2}{c}{50~au} & \multicolumn{2}{c}{200~au} \\
\cline{3-6}
 & & Total Number Particles & Flux (mJy km/s) & Total Number Particles & Flux (mJy km/s) \\
\hline
CS (6-5) & 1~G$_0$ & \(1.10 \times 10^{37}\) & 4.3 & \(2.40 \times 10^{37}\) & 46.1 \\
         & 10~G$_0$ & \(2.20 \times 10^{36}\) & 4.3 & \(2.20 \times 10^{37}\) & 35.0 \\
         & 50~G$_0$ & \(2.00 \times 10^{36}\) & 4.3 & \(2.10 \times 10^{37}\) & 24.3 \\
         & 100~G$_0$ & \(2.00 \times 10^{36}\) & 4.5 & \(2.10 \times 10^{37}\) & 20.9 \\
\hline
C (1-0)  & 1~G$_0$ & \(8.70 \times 10^{39}\) & 397.8 & \(1.60 \times 10^{41}\) & 933.8 \\
         & 10~G$_0$ & \(8.80 \times 10^{39}\) & 408.6 & \(1.80 \times 10^{41}\) & 1008.8 \\
         & 50~G$_0$ & \(9.10 \times 10^{39}\) & 508.3 & \(2.10 \times 10^{41}\) & 1157.1 \\
         & 100~G$_0$ & \(9.30 \times 10^{39}\) & 618.8 & \(2.10 \times 10^{41}\) & 1180.2 \\
\hline
HCO$^+$ (4-3) & 1~G$_0$ & \(2.40 \times 10^{37}\) & 2059.4 & \(1.20 \times 10^{38}\) & 6834.1 \\
              & 10~G$_0$ & \(2.30 \times 10^{37}\) & 1996.0 & \(1.00 \times 10^{38}\) & 6323.8 \\
              & 50~G$_0$ & \(2.20 \times 10^{37}\) & 2036.6 & \(1.00 \times 10^{38}\) & 6545.1 \\
              & 100~G$_0$ & \(2.10 \times 10^{37}\) & 2165.7 & \(1.10 \times 10^{38}\) & 6736.7 \\
\hline
N$_2$H$^+$ (3-2) & 1~G$_0$ & \(1.40 \times 10^{35}\) & 8.2 & \(1.90 \times 10^{36}\) & 70.1 \\
                 & 10~G$_0$ & \(1.40 \times 10^{35}\) & 7.7 & \(9.80 \times 10^{35}\) & 38.5 \\
                 & 50~G$_0$ & \(1.30 \times 10^{35}\) & 6.2 & \(6.60 \times 10^{35}\) & 25.7 \\
                 & 100~G$_0$ & \(1.30 \times 10^{35}\) & 5.2 & \(5.60 \times 10^{35}\) & 21.3 \\
\hline
C$_2$H (2-1) & 1~G$_0$ & \(1.90 \times 10^{38}\) & 0.9 & \(4.10 \times 10^{38}\) & 2.9 \\
             & 10~G$_0$ & \(1.90 \times 10^{38}\) & 0.8 & \(4.20 \times 10^{38}\) & 5.3 \\
             & 50~G$_0$ & \(1.90 \times 10^{38}\) & 0.8 & \(4.90 \times 10^{38}\) & 3.3 \\
             & 100~G$_0$ & \(1.90 \times 10^{38}\) & 0.8 & \(5.50 \times 10^{38}\) & 3.7 \\
\hline
CN (2-1) & 1~G$_0$ & \(4.10 \times 10^{37}\) & 5.1 & \(2.90 \times 10^{38}\) & 103.5 \\
         & 10~G$_0$ & \(4.00 \times 10^{37}\) & 5.0 & \(2.80 \times 10^{38}\) & 98.9 \\
         & 50~G$_0$ & \(4.00 \times 10^{37}\) & 4.9 & \(3.00 \times 10^{38}\) & 100.2 \\
         & 100~G$_0$ & \(3.90 \times 10^{37}\) & 4.8 & \(3.10 \times 10^{38}\) & 96.7 \\
\hline
HCN (4-3) & 1~G$_0$ & \(5.20 \times 10^{40}\) & 17.7 & \(6.10 \times 10^{40}\) & 188.5 \\
          & 10~G$_0$ & \(5.10 \times 10^{40}\) & 24.1 & \(6.10 \times 10^{40}\) & 219.9 \\
          & 50~G$_0$ & \(6.20 \times 10^{40}\) & 17.4 & \(6.10 \times 10^{40}\) & 88.9 \\
          & 100~G$_0$ & \(7.30 \times 10^{40}\) & 18.1 & \(6.10 \times 10^{40}\) & 79.7 \\
\hline
\end{tabular}
\end{table*}

The molecules discussed in Sections~\ref{sec:resultsneutrals} and \ref{sec:resultsions} have a variety of different responses to the magnitude of the ISRF impinging on the disk. Some species' column densities are not impacted at all while others vary by an order of magnitude. Local abundances per total hydrogen can change by even larger amounts. How will these changes manifest in observable signatures, and what can we learn from their spectra? To address this question, we post process select species from the chemical models using the ``Line Modeling Engine'' LIME, a non-LTE excitation and radiative transfer code \citep{brinch2010}. 
These simulations are produced utilizing collision rates from LAMDA database \citep{schoier2005}. We present spectra for the most UV sensitive species, neutral C, \hcop\, CS, and \nthp, using representative transitions for each. For the \nthp\ simulations, we utilized the collisional rates without hyperfine splitting from \citet{schoier2005}, which are based on HCO$^+$. The LIME models are then summed over to get the disk-integrated spectra, without added noise besides that introduced in the radiative transfer. 

Figure \ref{spectra} presents the spectra for our representative transitions, [CI] $J=1-0$, \hcop\ $J=4-3$, CS $J=6-5$, \nthp\ $J=3-2$, \cth\ $J=2-1$, CN $J=2-1$, and HCN $J=4-3$ for the 200~au and 50~au disk models at a distance of 100~pc. The trends with changing \GO\ tend to follow with the column density and abundance of these species. The peak flux density of [CI] $J=1-0$ in the 200~au model for 1~\GO\ is $\sim$220~mJy. As the \GO\ value increases to 100, the flux increases to 250~mJy. This follows the trend seen from the column density, as CO is dissociated, there is an increase in the column density of C, resulting in brighter emission. Interestingly, we see this same trend in the 50~au disk. Where this disk model was too compact to show much effect from the increasing UV field from other species, the peak flux of C increases in the outer disk as the external radiation field is increased. Therefore, C could potentially be a useful diagnostic of the effects of external irradiation in the compact disks seen in the ONC.

The line profile for \hcop\ is the brightest of these species, with a peak flux of 1400~mJy for the 200~au disk in purple. We see the slightest increase in the 100~\GO\ model, but otherwise they seem indistinguishable. The 50~au disk is fainter, with a flux of 400~mJy, but also does not show variation to the changing UV field. This insensitivity appears to be related to the line optical depth.

The flux density of CS decreases by a factor of 6 from 1~\GO\ to 100~\GO\ in the 200~au model. The peak flux of the 1~\GO\ model is $\sim$10 mJy. The 50~au model is extremely faint, with the peak flux not even 1 mJy. We do not observe the same trend of decreasing flux corresponding with increasing \GO\ in the 50~au disk. However, the 50 au disk model is extremely dense in comparison and more resistant to the effects of external UV. 

The flux of \nthp\ changes as the background FUV field is increased for both the 200~au and 50~au disk. For the 200~au disk, the flux appears to decrease by a factor of approximately two for each increasing step in the UV field, with a peak flux of $\sim$ 17.5 mJy for the 1~\GO\ model. The 50~au disk is about $10\times$ fainter. 

As the flux of C scales up with an increased ISRF, and the flux of \nthp\ and CS becomes more suppressed, a viable technique could be using ratios to detect changes in ionization in irradiated disks. 
The ratio of C/\nthp\ for the 200~au disk model for 1~\GO\ is $\sim$ 12.5, and in the 100~\GO\ model that ratio becomes 50. For the compact disk case the C/\nthp\ ratio for 1~\GO\ is 45 and that ratio increases for 100~\GO\ model to 140. The difference between C and CS is even more pronounced. As many of the irradiated disks in the ONC surveys have been shown to be compact and dense \citep{eisner2018}, this technique of using a ratio of species could be useful in detecting ionization trends or signatures in the outer disk.

The spectra for \cth\ $J=2-1$ shows extremely faint emission for both disk models, with the 50~au model showing a much lower flux than the 200~au model, consistent with the other species' as a result of the compact disk structure. The faint emission from \cth\ could be attributed to a lower C/O ratio in the models, as described in earlier sections, and varying this ratio could produce a more observable spectra. This is beyond the scope of this work, but worth further consideration given results from \citet{Boyden2023} on low degrees of CO sequestration in Orion disks. The broadening and double-peaked feature of \cth\ in the 50~au model is from the hyperfine components included in the model.

The simulated spectra for CN $J=2-1$, and HCN $J=4-3$ for the 200 and 50~au models is also shown in Fig \ref{spectra}. The external irradiation appears to have little effect on the CN line emission in both the 200~au model and the 50~au model, as we do not see any change between the four ISRF models. The 50~au model follows the trend of decreased flux, but we also see a broadening that is not present in the 200~au model for both CN and HCN. HCN is also much more affected by the external UV in both the 50 and 200~au disk models. Interestingly, the flux for HCN increases from 1~\GO\ to 10~\GO, and then rapidly decreases at 50 and 100~\GO. Results from \citet{walsh2013} for HCN $J=4-3$, shows a marked increase in the flux in the irradiated model from the isolated model compared to our models. Note, however, that the irradiated model in \citet{walsh2013} is $4 \times 10^{5}$~\GO, an extremely high UV field, compared to our moderate \GO\ levels. As stated above in the sections regarding CN and HCN, we do not account for vibrationally excited \hh, which is a formation pathway for HCN important especially in irradiated environments. CN also shows an increase in flux in \citet{walsh2013} between the isolated and irradiated models, whereas in our models we show little change between the low to moderately irradiated models.

Although not shown, the maximum flux of CO $J=3-2$ in the 200 au 1~\GO\ model is around 3~Jy and so the CO/\nthp\ ratio at 1~\GO\ is $\sim170$. Increasing the ISRF to 100~\GO, results in a peak CO line flux of  4~Jy. Taken with the \nthp\, the resulting CO/\nthp\ ratio is 1600, which would constitute a $\sim 8.36 \times$ increase.

Note, all of the models presented assume a default distance of 100~pc. While the ratios will not be impacted, the absolute line fluxes should be scaled to the distance of a given cluster like the ONC. Fainter emission from species like \nthp\ will be challenging, even for ALMA. \hcop, with a much brighter flux, is still a good candidates for testing irradiated disk chemistry. 
Indeed, \citet{Boyden2023} found that \hcop\ emission in the ONC proplyds is very bright and was additionally less impacted by contamination from the background cloud than, e.g., CO.

\subsection{Inner disk vs. outer disk chemistry}

An interesting feature of all of our chemical models is that the inner disk region, i.e., within $\sim$25 au, is largely similar regardless of \GO\ value. There are some differences in the the surface abundance, e.g., of CS at $Z/R=0.2$  (Figure~\ref{abund_neutrals}), but these do not appear to impact the overall column density significantly. The midplane is essentially identical, which is consistent with the high optical depths to UV photons. Instead, we find that the ionization budget of the inner disk is indeed governed by the UV and X-ray fields from the central star, which dominates any contribution from the external environment at these moderately enhanced external UV fields. 

This feature of our models is also reflected in the spectra (Figure~\ref{spectra}), where for all species, the line profiles primarily vary at low velocities close to the systemic velocity (0~km~s$^{-1}$ here). These velocities correspond to the outer disk. The high velocity wings that trace the inner disk gas remain largely unchanged regardless of \GO. 
Thus, future high spectral resolution observations of UV-sensitive tracers will be able to use a combination of the line profiles and molecular ratios (Section~\ref{sec:LIMEsim}) to confirm the degree to which disk chemistry is altered by the cluster environment on scales relevant to planet formation, closer to the central star. 

Moreover, we find that even the smallest, 50~au outer radius disk model is not impacted strongly at $<25$~au. Since this physical structure is relatively dense, it efficiently shields the disk midplane at almost all radii. Our solar system was thought to be a small, similarly sized protoplanetary disk in an Orion-like environment \citep[e.g.,][]{kretke2012}. Perhaps its small size and relatively large mass provided a similar shielding effect, such that it not only could survive photoevaporation but may have also shielded its molecular content from the external UV environment.

More generally, these inner disk results imply that we can learn a substantial amount about the chemistry of planet formation at $<25$~au scales even from disks in less extreme environments such as Taurus or Lupus, which are significantly closer and easier to observe than the more distant disks of the ONC. This finding holds specifically for the UV ISRF, however, and it is possible that other sources of external irradiation like externally enhanced X-rays or cosmic rays \citep{adams2010} may contribute in different ways depending on the details of the cluster environment.

\section{Summary}\label{sec:conclusions}
We present a case study of the chemistry in moderately irradiated protoplanetary disks exposed to \GO\ values of 1 to 100. We explore how varying the external UV ISRF changes the disk chemistry for disks of varying outer radius, but for a fixed mass of 0.01 M\textsubscript{\(\odot\)}. Together, this suite of models allows us to study the interplay of the disk structure and the external UV in setting the bulk molecular content of the disk. We specifically present abundances of molecules that have either been observed in disks (e.g., HCO$^+$) or have otherwise impactful chemistry (e.g., C$^+$). We find that: 

\begin{itemize} 

        \item The species most affected by the increase in external UV are \nthp, \cp, CS, C, and HCN. Both the increased column density and abundance for C and \cp\ are caused by the increased photodesorption and photodissociation of CO in the outer disk from increased UV photons. The \nthp\ abundance and column density is fundamentally changed in an enhanced UV field, decreasing as the external UV increases. The latter will complicate the interpretation of \nthp\ as a snow line tracer in externally irradiated disks.

        \item The CN/HCN column density is enhanced as the external UV is increased, which is an expected result based on observations and results from previous modeling studies. As UV flux increases, there will be increased photodissociation of HCN in the surface and outer radii of the disk. This effect is seen to the greatest extent in the 200~au model, where UV photons can penetrate deeper into the disk and affect the chemistry.
        
        \item While the column density of CO is not strongly changed with increasing external UV, the midplane abundance is altered. In the 200~au disk model there appears to be a secondary CO snowline at the outer edge of the irradiated disk models due to enhanced photodesorption of CO ice. 
        
        \item Not all species are monotonically impacted -- positively or negatively -- by increasing \GO. For example, the column density of \hcop\ in the 100~au and 200~au outer radius models at first decreases and then increases again as the UV flux increases, depending on radial location. This behavior relates to the chemistry of its precursor, CO, and one of its key destructive reactions, recombination with electrons.

        \item The density structure of the disk plays an important role in how the external UV makes its way into the disk. For example, the 50~au disk model is very dense and compact and therefore helps shield the disk from the enhanced UV field, even at the outermost edges. The more extended 200~au model allows more UV photons to penetrate and impact the outer disk chemistry, particularly affecting the photosensitive species like C, \nthp, \cp, and CS. 

        \item The inner radial 25~au is relatively unchanged in all three disk models by increasing \GO. This has important implications for the evolution of the chemistry in the ``terrestrial planet forming zone'' -- i.e., the inner 10~au. One key implication is that we should be able to compare the chemistry of disks in clusters with massive stars to those in lower mass star forming regions, which are relatively easier to study. 

        \item We examine synthetic spectra for a subset of the molecules. The emission from neutral [CI] $J=1-0$ and \hcop\ $J=4-3$ had the brightest overall flux densities of the species modeled. For \hcop, the competing chemical effects combined with line optical depth will make it challenging to use as a UV field diagnostic. 
        Neutral C emission is a promising tracer of ionization in the outer disk in both the compact and extended disk models. The ratio of C/\nthp\ could also be used as a potential external UV field tracer, given that \nthp\ decreases with increasing UV. Unfortunately, \nthp\ is quite faint in all of our models, likely requiring extensive integration times even with ALMA at a distance of the ONC. Alternatively, ratios of C/CS could also be used to constrain the external UV field, as CS follows the same decreasing trend with \GO\ as \nthp.
    \end{itemize}  

As ALMA begins more extensive surveys of disk chemistry for stars in cluster environments, it will be important to understand the diverse kinds of environments present, from the cluster center to the outer edges. We find that the outer reaches of the cluster sets up a distinct set of conditions from the more highly irradiated disks near massive stars, typically at the cluster center. While the outer edges of the disk are still impacted, the innermost radii do not show strongly altered chemistry, though perhaps additional effects, like radial drift of solids, may link these environments. Differences in these radial zones will be important to consider, since most clusters are substantially further, hindering our ability to spatially resolve the chemistry as easily as nearby disks. Thus alternative tools, including using Doppler motion where possible will be important for disentangling the different radial environments. 

\begin{acknowledgments}
REG acknowledges support from the University of Virginia Interdisciplinary Doctoral Fellowship and the Johnson \& Johnson WiSTEM2D program. LIC acknowledges support from the David and Lucille Packard Foundation, Research Corporation for Science Advancement Cottrell Fellowship, NASA ATP 80NSSC20K0529, and NSF grant no. AST-2205698. 
\end{acknowledgments}

\bibliographystyle{aasjournal}

\appendix

\section{Neutral abundances for 100 au and 50 au disk models}
\begin{figure}[H]
    \centering
    \includegraphics[scale=0.175]{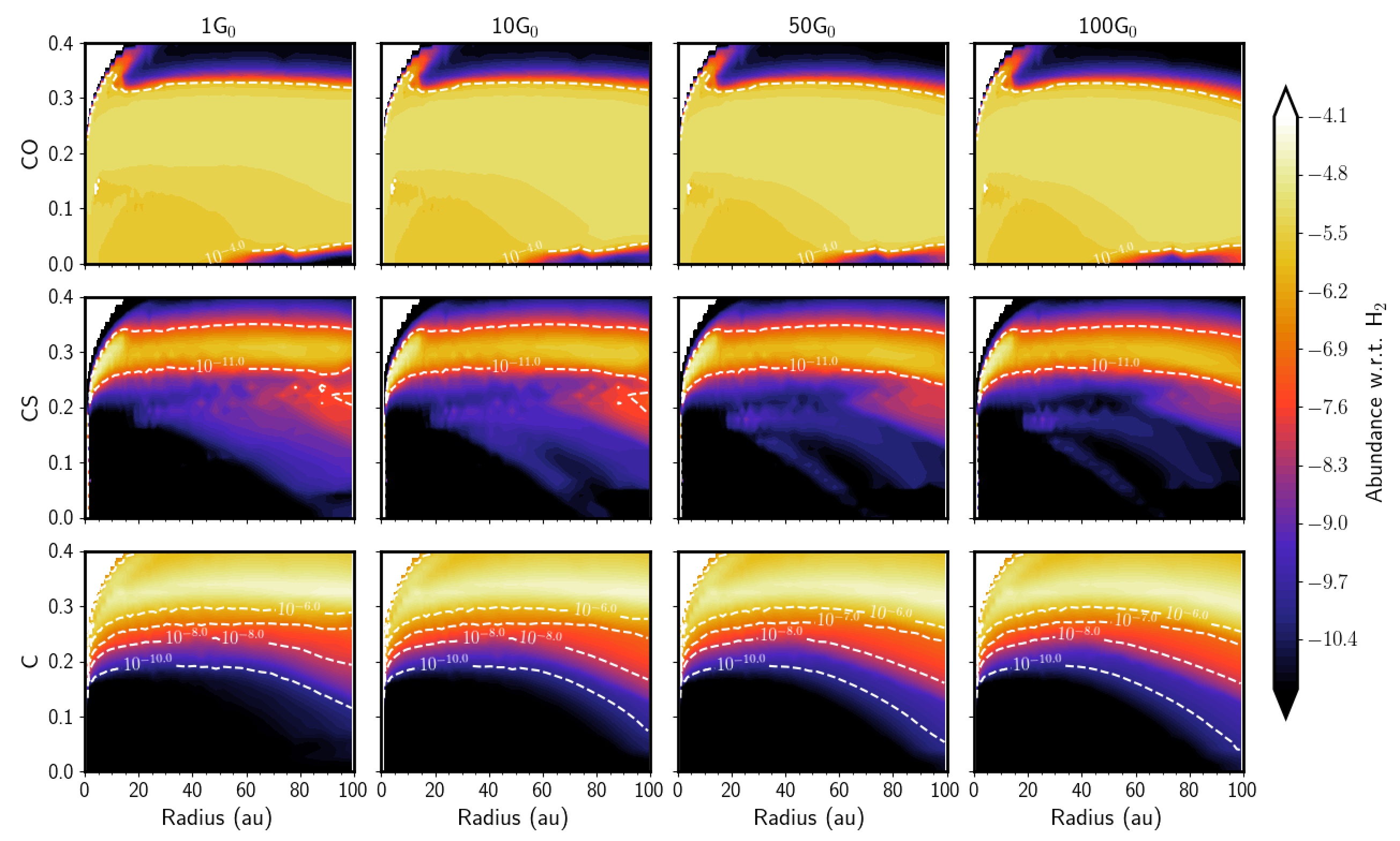}
    \caption{2D contour plots of CO, CS, and C abundances with respect to \hh\ for the 100~au disk at varying external UV fields.} 
    
    \label{abund_neutrals100}
\end{figure}

\begin{figure}[H]
    \centering
    \includegraphics[scale=0.175]{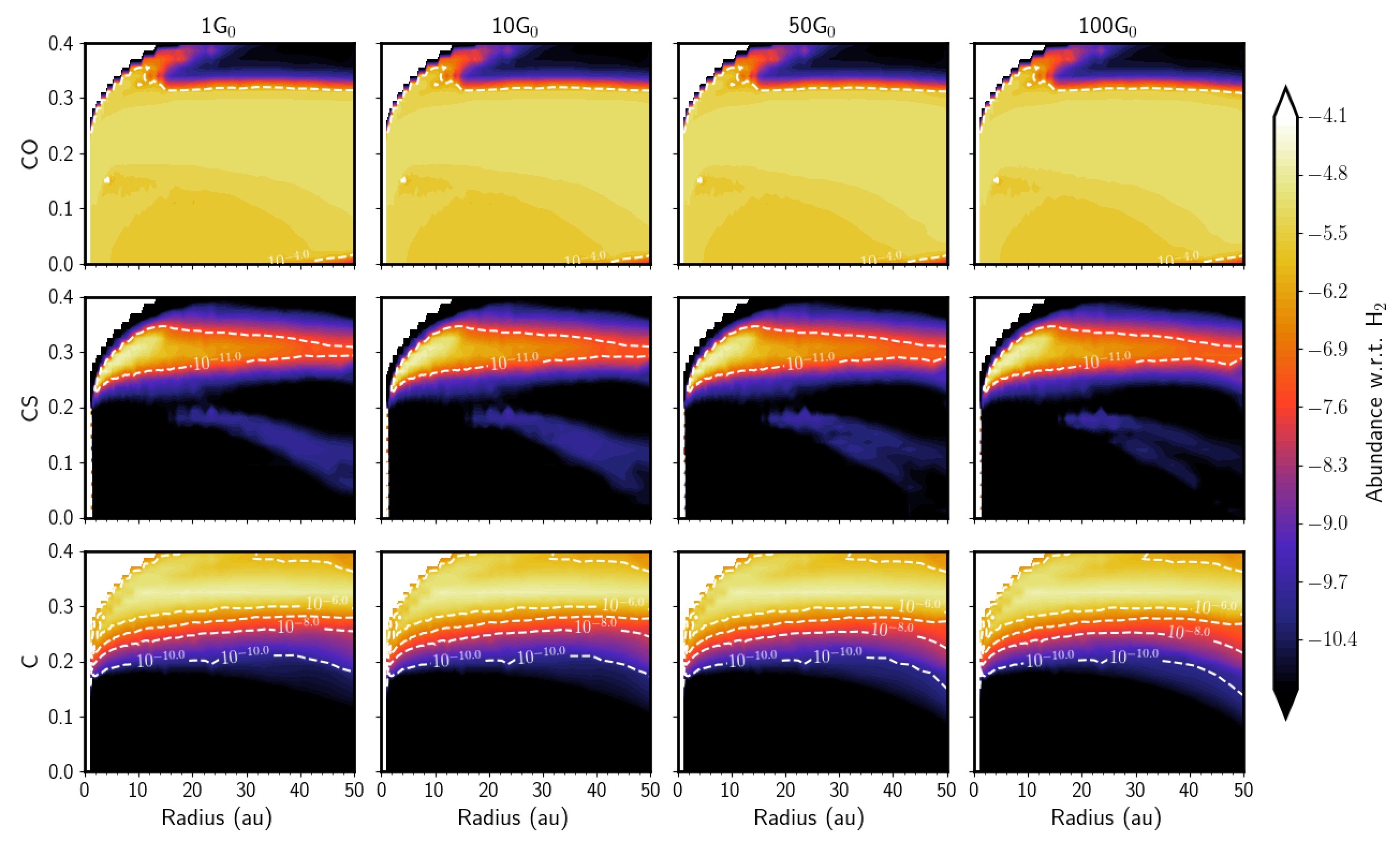}
    \caption{2D contour plots of CO, CS, and C abundances w.r.t \hh\ for the 50~au disk at varying external UV fields.} 
    
    \label{abund_neutrals50}
\end{figure}

\begin{figure}[H]
    \centering
    \includegraphics[scale=0.175]{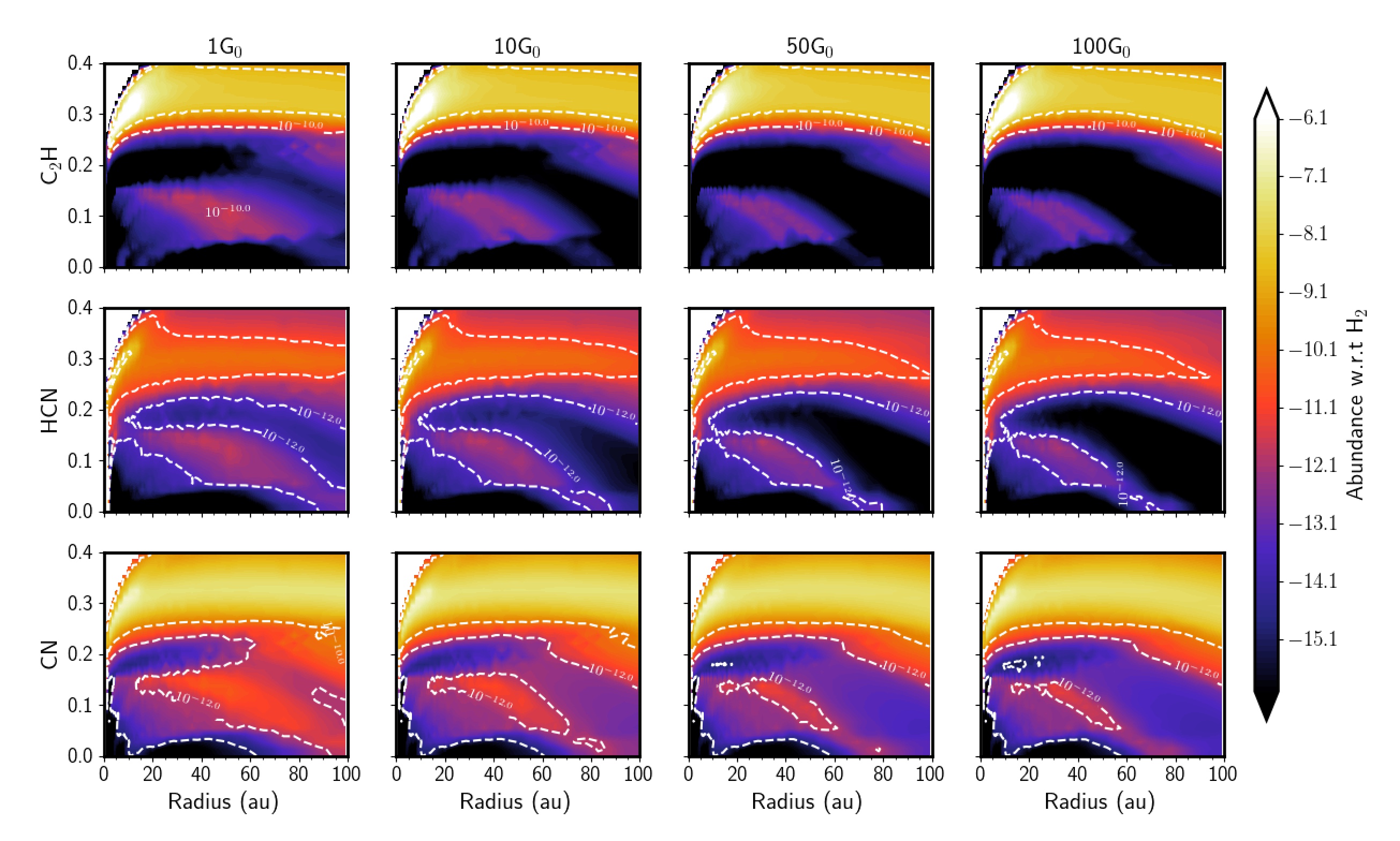}
    \caption{2D contour plots of \cth, HCN, and CN abundances with respect to \hh\ for the 100~au disk at varying external UV fields.} 
    
    \label{hcn_neutrals100}
\end{figure}

\begin{figure}[H]
    \centering
    \includegraphics[scale=0.175]{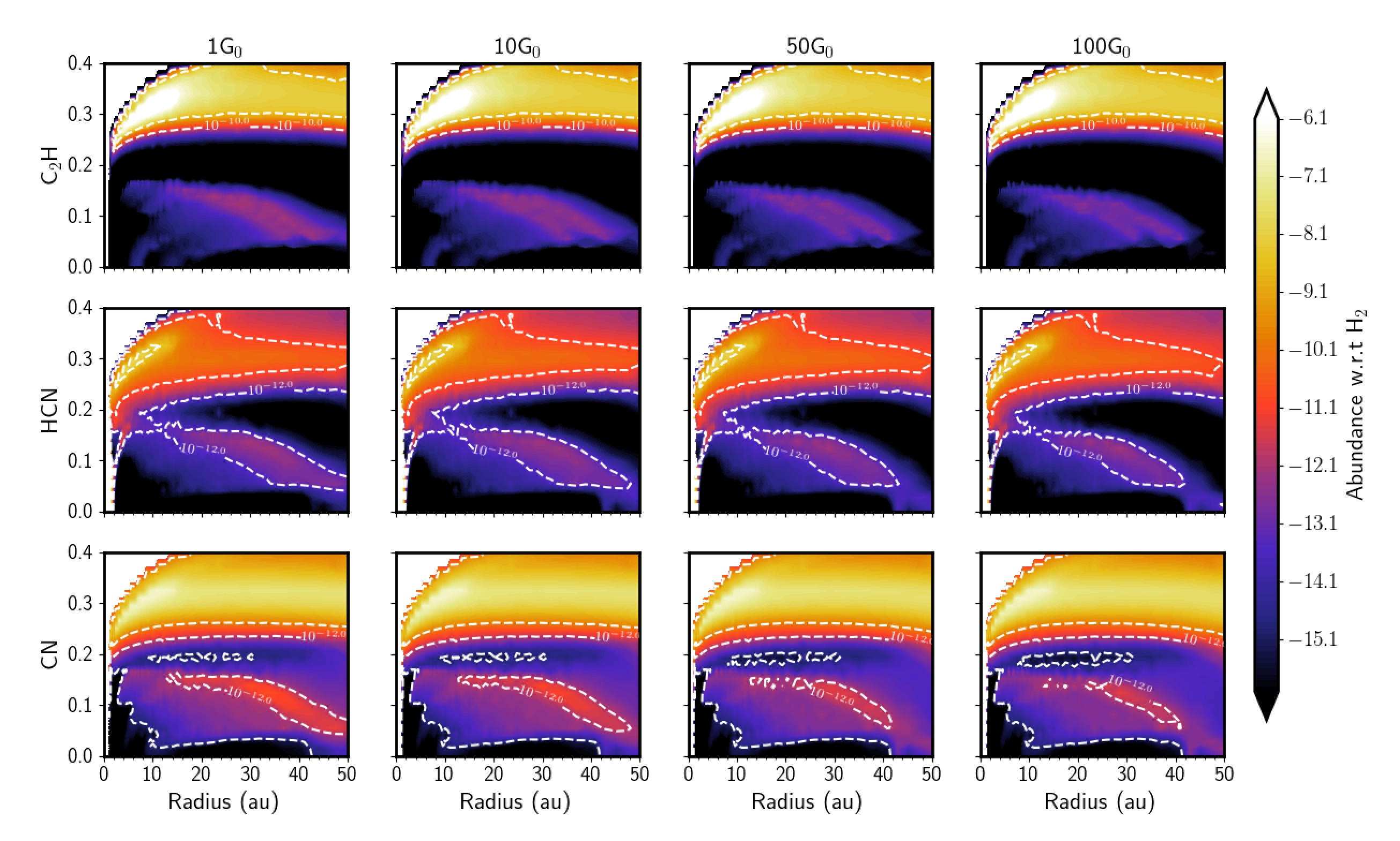}
    \caption{2D contour plots of \cth, HCN, and CN abundances w.r.t \hh\ for the 50~au disk at varying external UV fields.} 
    
    \label{hcn_neutrals50}
\end{figure}

Figures~\ref{abund_neutrals100}, \ref{abund_neutrals50}, \ref{hcn_neutrals100}, and \ref{hcn_neutrals50} present the 2D abundance distribution of the neutral species CO, CS, C, \cth, CN, and HCN for the 100 and 50~au outer radius disk models, respectively. Compared to the 200~au model, they change less with increasing external UV field. As both of these models are denser than the more extended 200~au model, they are more impervious to the effects of external ionization from the UV source. 

\section{Ion abundances for 100~au and 50~au disk models}
\begin{figure}[H]
    \centering
    \includegraphics[scale=0.175]{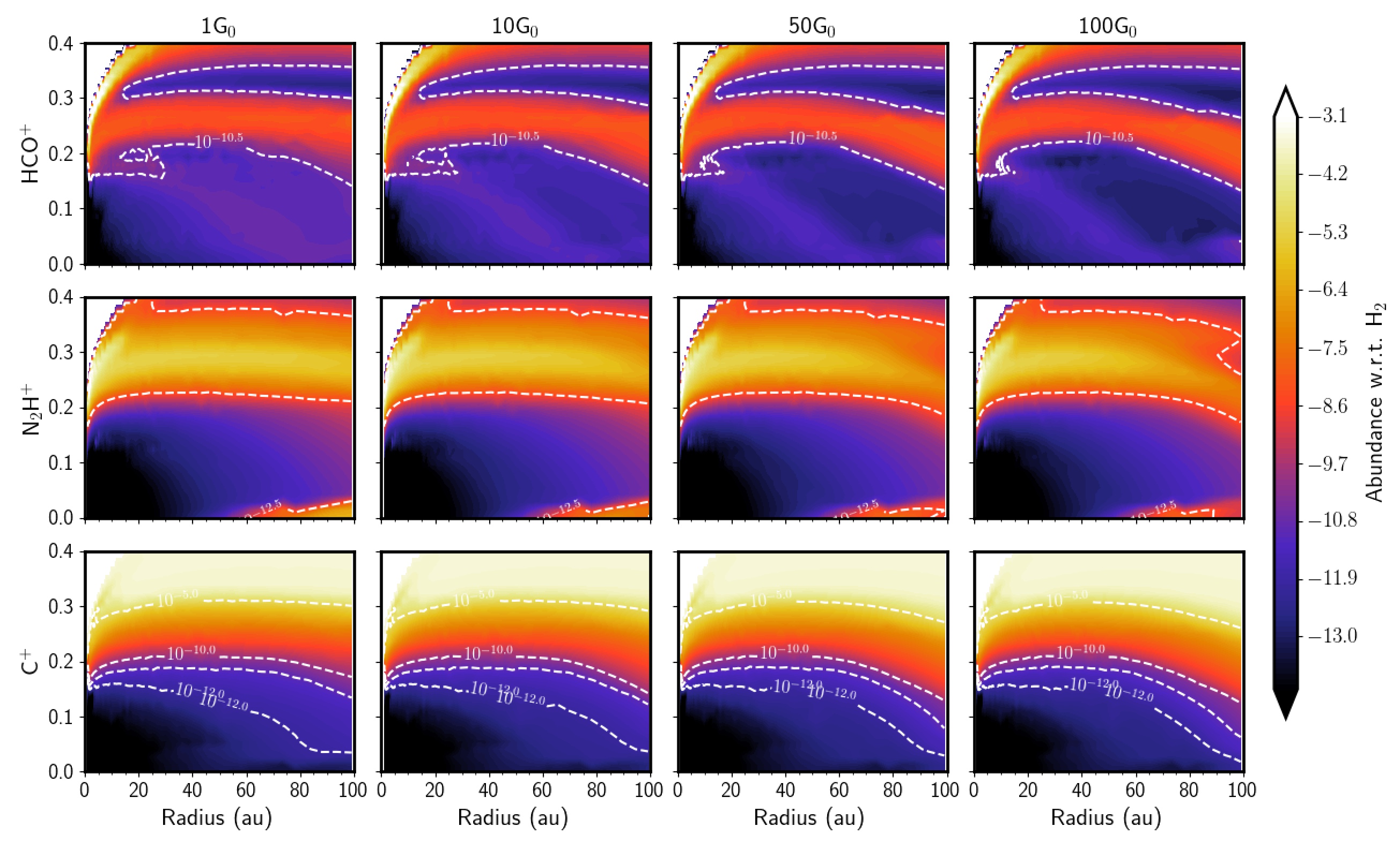}
    \caption{2D contour plots of \hcop, \nthp, and \cp\ abundances with respect to \hh\ for the 100 au disk at varying external UV fields.} 
    
    \label{ions_abund100}
\end{figure}

\begin{figure}[H]
    \centering
    \includegraphics[scale=0.175]{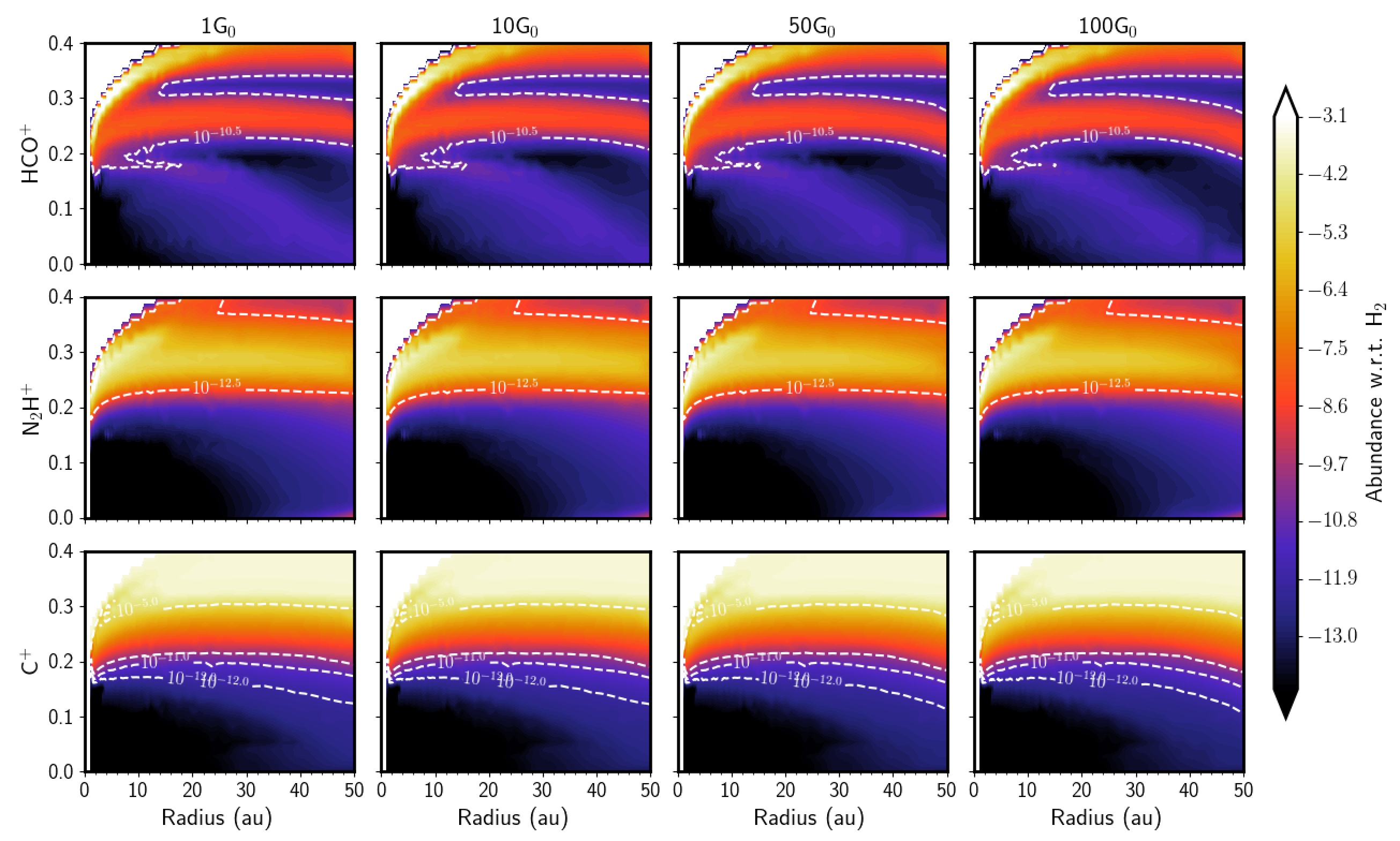}
    \caption{2D contour plots of \hcop, \nthp, and \cp\ abundances with respect to \hh\ for the 50 au disk at varying external UV fields.} 
    
    \label{ions_abund50}
\end{figure}

Figures~\ref{ions_abund100} and \ref{ions_abund50} present the abundances for the ions \nthp, \hcop, and \cp\ for the 100 and 50~au disk models, respectively. Similar to the neutral case in Appendix A, the denser disk models do not show as strong of an effect from increased external UV compared to the 200~au disk model.

\end{document}